\documentclass[preprint]{aastex}

\shorttitle{Ultra-compact structure in intermediate-luminosity radio
quasars} \shortauthors{Cao et al.}

\begin{document}


\title{Ultra-compact structure in intermediate-luminosity radio quasars: building a sample of standard cosmological rulers and improving the dark energy constraints up to $z\sim 3$}


\author{Shuo Cao}
\affil{Department of Astronomy, Beijing Normal University,
    Beijing 100875, China}

\author{Xiaogang Zheng}
\affil{Department of Astronomy, Beijing Normal University,
    Beijing 100875, China; \\
    Department of Astrophysics and Cosmology, Institute of Physics, University of Silesia, Uniwersytecka 4, 40-007, Katowice, Poland}

\author{Marek Biesiada}
\affil{Department of Astronomy, Beijing Normal University,
    Beijing 100875, China; \\
    Department of Astrophysics and Cosmology, Institute of Physics, University of Silesia, Uniwersytecka 4, 40-007, Katowice, Poland}

\author{Jingzhao Qi}
\affil{Department of Astronomy, Beijing Normal University,
    Beijing 100875, China}

\author{Yun Chen}
\affil{Key Laboratory for Computational Astrophysics, National
Astronomical Observatories, Chinese Academy of Sciences, Beijing,
100012, China}

\and

\author{Zong-Hong Zhu}
\affil{Department of Astronomy, Beijing Normal University,
    Beijing 100875, China}
\email{zhuzh@bnu.edu.cn}



\begin{abstract}

In this paper, we present a new compiled milliarcsecond compact
radio data set of 120 intermediate-luminosity quasars in the
redshift range $0.46< z <2.76$. These quasars show negligible
dependence on redshifts and intrinsic luminosity, and thus
represents, in the standard model of cosmology, a fixed
comoving-length of standard ruler. We implement a new
cosmology-independent technique to calibrate the linear size of of
this standard ruler as $l_m= 11.03\pm0.25$ pc, which is the typical
radius at which AGN jets become opaque at the observed frequency
$\nu\sim 2$ GHz. In the framework of flat $\Lambda$CDM model, we
find a high value of the matter density parameter,
$\Omega_m=0.322^{+0.244}_{-0.141}$, and a low value of the Hubble
constant, $H_0=67.6^{+7.8}_{-7.4}\; \rm{kms}^{-1}\rm{Mpc}^{-1}$,
which is in excellent agreement with the CMB anisotropy measurements
by \textit{Planck}. We obtain ${\Omega_m}=0.309^{+0.215}_{-0.151}$,
$w=-0.970^{+0.500}_{-1.730}$ at 68.3\% CL for the constant $w$ of a
dynamical dark-energy model, which demonstrates no significant
deviation from the concordance $\Lambda$CDM model. Consistent
fitting results are also obtained for other cosmological models
explaining the cosmic acceleration, like Ricci dark energy (RDE) or
Dvali-Gabadadze-Porrati (DGP) brane-world scenario. While no
significant change in $w$ with redshift is detected, there is still
considerable room for evolution in $w$ and the transition redshift
at which $w$ departing from -1 is located at $z\sim 2.0$. Our
results demonstrate that the method extensively investigated in our
work on observational radio quasar data can be used to effectively
derive cosmological information. Finally, we find the combination of
high-redshift quasars and low-redshift clusters may provide an
important source of angular diameter distances, considering the
redshift coverage of these two astrophysical probes.

\end{abstract}





\section{Introduction} \label{intro}

That the expansion of the Universe is accelerating at the current
epoch has been demonstrated by the observations of Type Ia
supernovae (SN Ia) \citep{Riess98, Perlmutter99} and also supported
by other independent probes, such as the Cosmic Microwave Background
(CMB) \citep{Pope04} and the Large Scale Structure (LSS)
\citep{Spergel03}. Therefore, the so called dark energy (DE), a new
component driving the observed accelerated expansion of the Universe
was introduced into the framework of general relativity. However,
the nature of this exotic source with negative pressure has remained
an enigma.

Besides the cosmological constant $\Lambda$ \citep{Peebles03}, the
simplest candidate consistent with current observations, which
however suffers from the well-known fine tuning and coincidence
problems, other possible dark energy models with different dark
energy equation of state (EoS) parametrizations
\citep{Ratra98,Chevalier01,Linder03} have also been the focus of
investigations in recent decades. Meanwhile, it should be noted that
cosmic acceleration might also be explained by possible departures
of the true theory of gravity from General Relativity. From these
theoretical motivations, possible multidimensionality in the brane
theory gave birth to the well-known Dvali-Gabadadze-Porrati (DGP)
model, while the holographic principle has generated the Ricci dark
energy (RDE) model. All above mentioned models are in agreement with
some sets of observational data, e.g. the distance modulus from
SNIa, or the CMB anisotropies. At this point, it is worth to
highlight the issue of evolving the equation of state $w(z)$ of dark
energy. Namely, a dynamical DE will be indicated by $w(z)$ evolving
across -1, rather than having a fixed value $w(z)=-1$, which implies
an additional intrinsic degree of freedom of dark energy, and could
be a smoking gun of the breakdown of Einstein's theory of general
relativity on cosmological scales \citep{Zhao12}. However, due to
the so-called ``redshift desert'' \footnote{SN Ia are commonly
accepted standard candles in the Universe and from their observed
distance moduli we are able to recover luminosity distances covering
the lower redshift range $z\leq 1.40$. On the other side, CMB
measurements, e.g. the latest results from \textit{Planck} probe
very high redshift $z\sim 1000$ corresponding to the last scattering
surface. Therefore the redshift range $1.40\leq z \leq 1000$ is
sometimes called the ``redshift desert'', because of fundamental
difficulties in obtaining observational data in this range.}, it is
very difficult to check dynamical DE from astrophysical
observations. When confronted with such theoretical and
observational puzzles, we have no alterative but turn to
high-precision data and develop new complementary cosmological
probes at higher redshifts. In this paper we propose that compact
structure measurements in radio quasars leading to calibrated
standard rulers can become a useful tool for differentiating between
the above mentioned dark energy models and exploring possible
dynamical evolution of $w(z)$.

In the framework of standard cosmology, over the past decades
considerable advances have been made in the search for possible
candidates to serve as ``true'' standard rods in the Universe. In
particular, cosmological tests based on the angular size ---
distance relation have been developed in a series of papers, and
implemented using various astrophysical sources. Recently, attention
of has been focused on large comoving length scales revealed in the
baryon acoustic oscillations (BAO). The BAO peak location is
commonly recognized as a fixed comoving ruler of about $105h^{-1}\;
Mpc$ (where $h$ is the Hubble constant $H_0$ expressed in units of
$100 \; km \, s^{-1} Mpc^{-1}$). However, the so-called fitting
problem \citep{Ellis87} still remains a challenge for BAO peak
location as a standard ruler. In particular, the environmental
dependence of the BAO location has recently been detected by
\citet{Roukema2015,Roukema2016}. Moreover, \citet{Ding2015} and
\citet{Zheng2016} pointed out a noticeable systematic difference
between $H(z)$ measurements based on BAO and those obtained with
differential aging techniques. Much efforts have also been made to
explore the sizes of galaxy clusters at different redshifts, by
using radio observations of the Sunyaev-Zeldovich effect together
with X-ray emission \citep{Filippis05,Bonamente06}. However, the
large observational uncertainties of these angular diameter distance
measurements significantly affect the constraining power of this
standard ruler. Actually, clusters alone could not provide a
competitive source of angular diameter distance to probe the
acceleration of the Universe.

In the similar spirit, radio sources constitute a specially powerful
population to test the redshift - angular size relation for extended
FRIIb galaxies \citep{Daly03}, radio galaxies
\citep{Guerra98,Guerra00}, and radio loud quasars
\citep{Buchalter98}. For instance, it was firstly proposed that the
canonical maximum lobe size of radio galaxies may provide a standard
ruler for cosmological studies. From the mean observed separation of
a sample of 14 radio lobes, in combination with the measurements of
radio lobe width, lobe propagation velocity, and inferred magnetic
field strength, \citet{Guerra98} found $\Omega_m =
0.2^{+0.3}_{-0.2}$ (68\% confidence) for a flat cosmology.

More promising candidates in this context are ultra-compact
structure in radio sources (especially for quasars that can be
observed up to very high redshifts), with milliarcsecond angular
sizes measured by very-long-baseline interferometry (VLBI)
\citep{Kellermann93,Gurvits94}. For each source, the angular size is
defined as the separation between the core (the strongest component)
and the most distant component with 2\% of the core peak brightness
\citep{Kellermann93}. The original data set compiled by
\citet{Gurvits99} comprises 330 milliarcsecond radio sources
covering a wide range of redshifts and includes various optical
counterparts, such as quasars and radio galaxies (hereafter we call
this data G99 for short). After excluding sources with synthesized
beam along the direction of apparent extension, compact sources
unresolved in the observed VLBI images were naturally obtained.
Finally, in order to minimize possible dependence of angular size on
spectral index and luminosity, the final sample was restricted to
145 compact sources with spectral index ($-0.38\leq \alpha\leq
0.18$) and total luminosity ($Lh^2 \geq 10^{26} \; WHz^{-1}$).
Possible cosmological application of these compact radio sources as
a standard rod has been extensively discussed in the literature
\citep{Vishwakarma01,Zhu02,Chen03}. In their analysis, the full data
set of 145 sources was distributed into twelve redshift bins with
about the same number of sources per bin. The lowest and highest
redshift bins were centered at redshifts $z=0.52$ and $z=3.6$
respectively. However, the typical value of the characteristic
linear size $l_m$ remained one of the major uncertainties in their
analysis. In order to provide tighter cosmological constraints, some
authors chose to fix $l$ at specific values
\citep{Vishwakarma01,Lima02,Zhu02}, while \citet{Chen03} chose to
include a large range of value for $l$ and then integrate it over to
obtain the probability distribution of parameters of interest.

The controversy around the exact value of the characteristic linear
size $l_m$ for this standard rod or even whether compact radio
sources are indeed ``true'' standard rods still existed. Under the
assumption of a homogeneous, isotropic universe without cosmological
constant, \citet{Gurvits99} and \citet{Vishwakarma01} suggested that
the exclusion of sources with extreme spectral indices and low
luminosities might alleviate the dependence of $l_m$ on the source
luminosity and redshift. More recently, \citet{Cao15} reexamined the
same data in the framework of $\Lambda$CDM cosmological model, and
demonstrated that both source redshift and luminosity will affect
the determination of the radio source size, i.e. the mixed
population of radio sources including different optical counterparts
(quasars, radio galaxies, etc.) cannot be treated as a  ``true''
standard rod. In their most recent work, however, by applying the
popular parametrization $l_m=lL^\beta(1+z)^n$, \citet{Cao16} found
that, compact structure in the intermediate-luminosity radio quasars
could serve as a standard cosmological rod minimizing the above two
effects ($|n|\simeq 10^{-3}$, $|\beta|\simeq 10^{-4}$), and thus
provide valuable sources of angular diameter distances at high
reshifts ($z\sim3.0$), reaching beyond feasible limits of supernova
studies. On the other hand, through the investigation of the
calibrated value for $l_m$, we will focus on the astrophysical
implication of the linear size for this standard ruler. As we
``look'' into the jet of AGN, the plasma is initially optically thin
(transparent), but gets less as we look further in and the plasma
density increases; eventually the plasma becomes optically thick
(opaque), which point we identify with the ``core''.

The focus of this paper is on two issues. First, we extend recent
analysis \citet{Cao16} of compact sources as standard rulers. The
extension is related to completely cosmological-model independent
calibration of the linear size of standard rulers. This was not
possible in previous study where the speed of light was discussed.
Then, the angular size measurements of 120 quasars covering redshift
range $z = 0.46 - 2.76$ will be used to constrain dynamical
properties of dark energy in a way competitive with other probes and
reaching to higher redshifts than other distance indicators. This
way we will demonstrate the usefulness of the sample presented.
The outline of the paper is as follows: in Section 2 we briefly
describe the quasar data and a cosmological model independent method
of calibrating the linear size of this standard ruler. In Section~3
we report the results of constraints on the properties of dark
energy obtained with the quasar data. Section 4 is devoted to
constraints on other mechanisms explaining the cosmic acceleration.
Finally, we give our discussion and conclusions in Section 5 and 6,
respectively.

\begin{figure*}
\begin{center}
\includegraphics[scale=0.85]{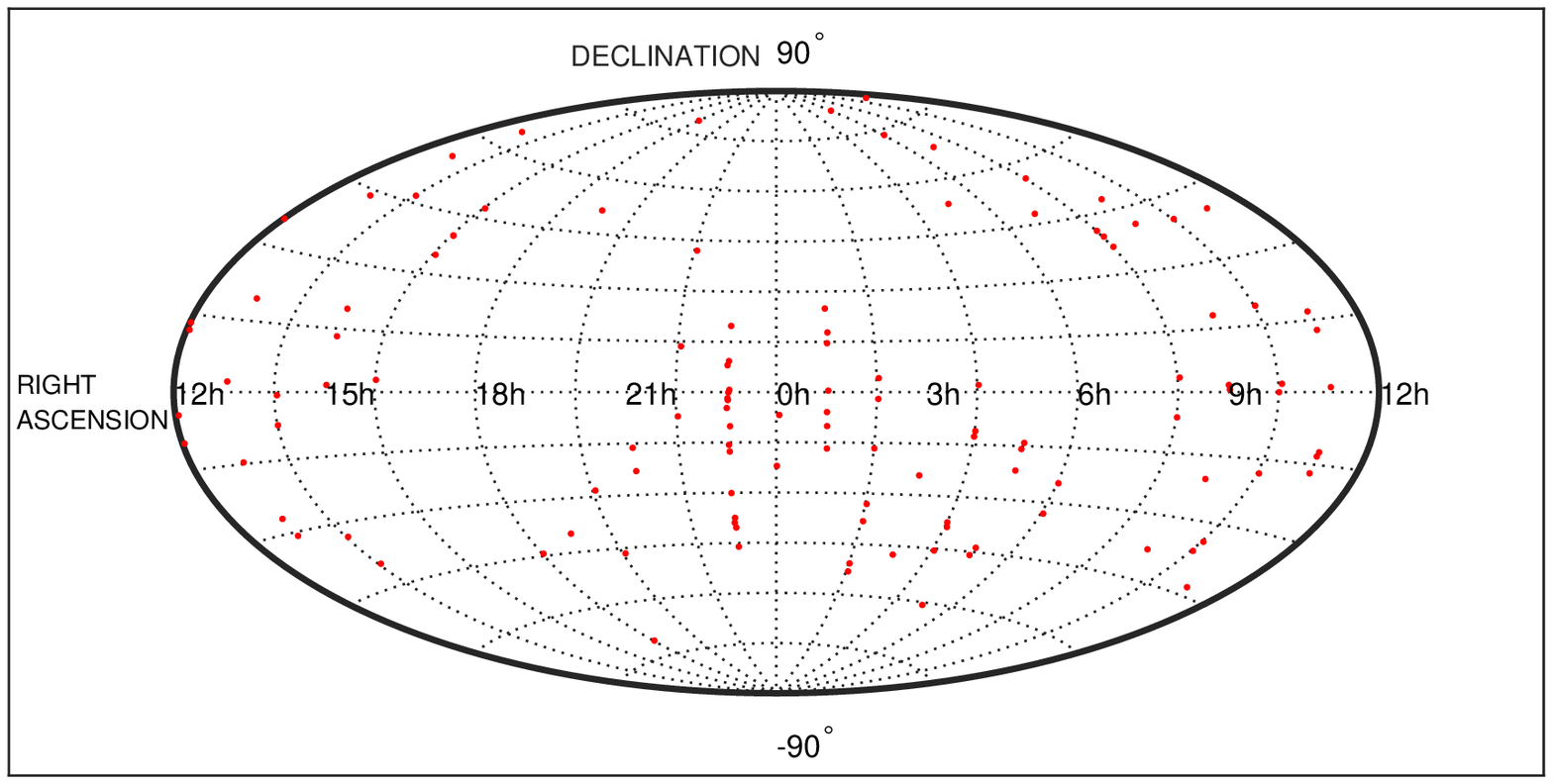} \\
\includegraphics[width=5cm,angle=0]{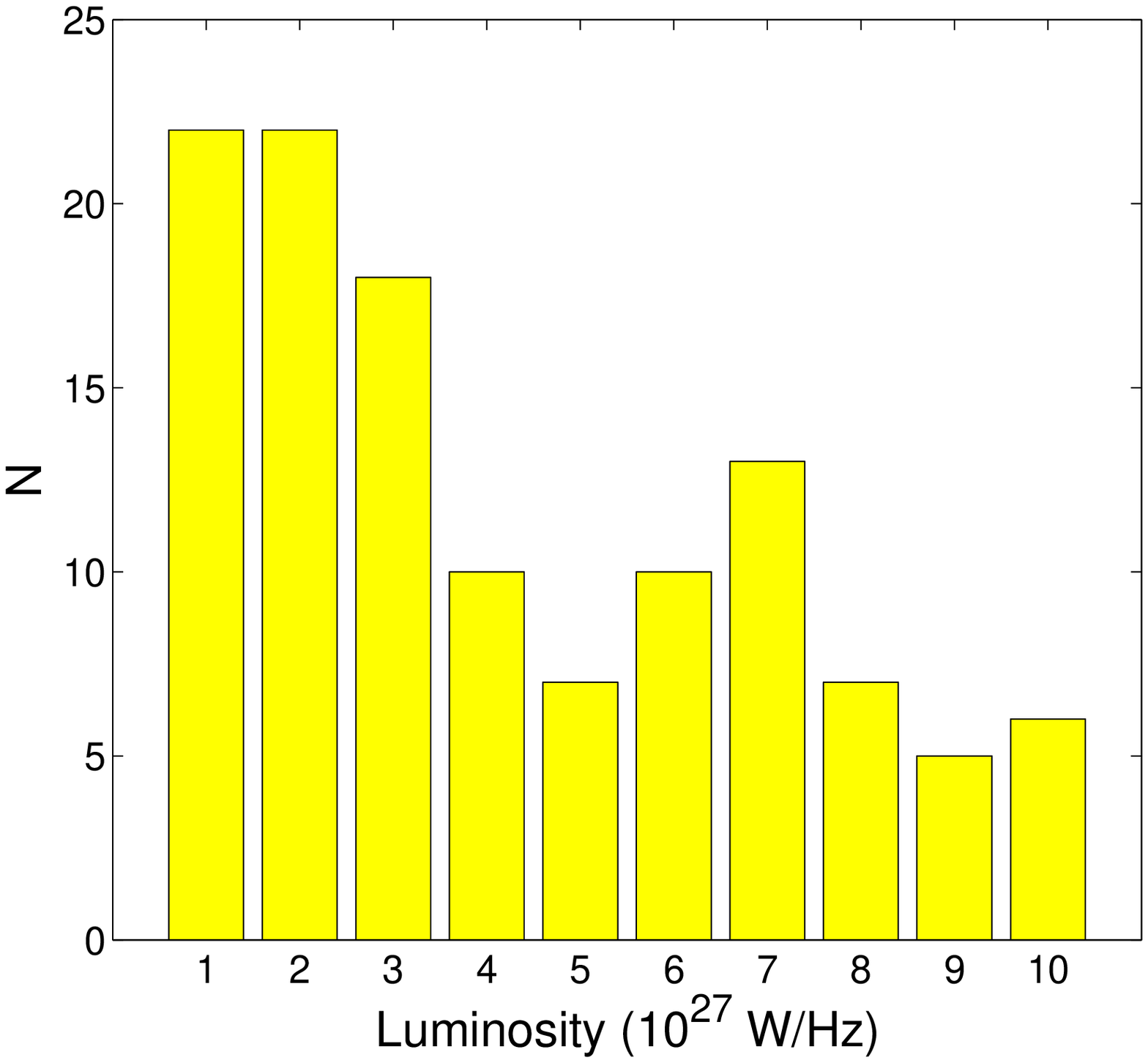} \includegraphics[width=5cm,angle=0]{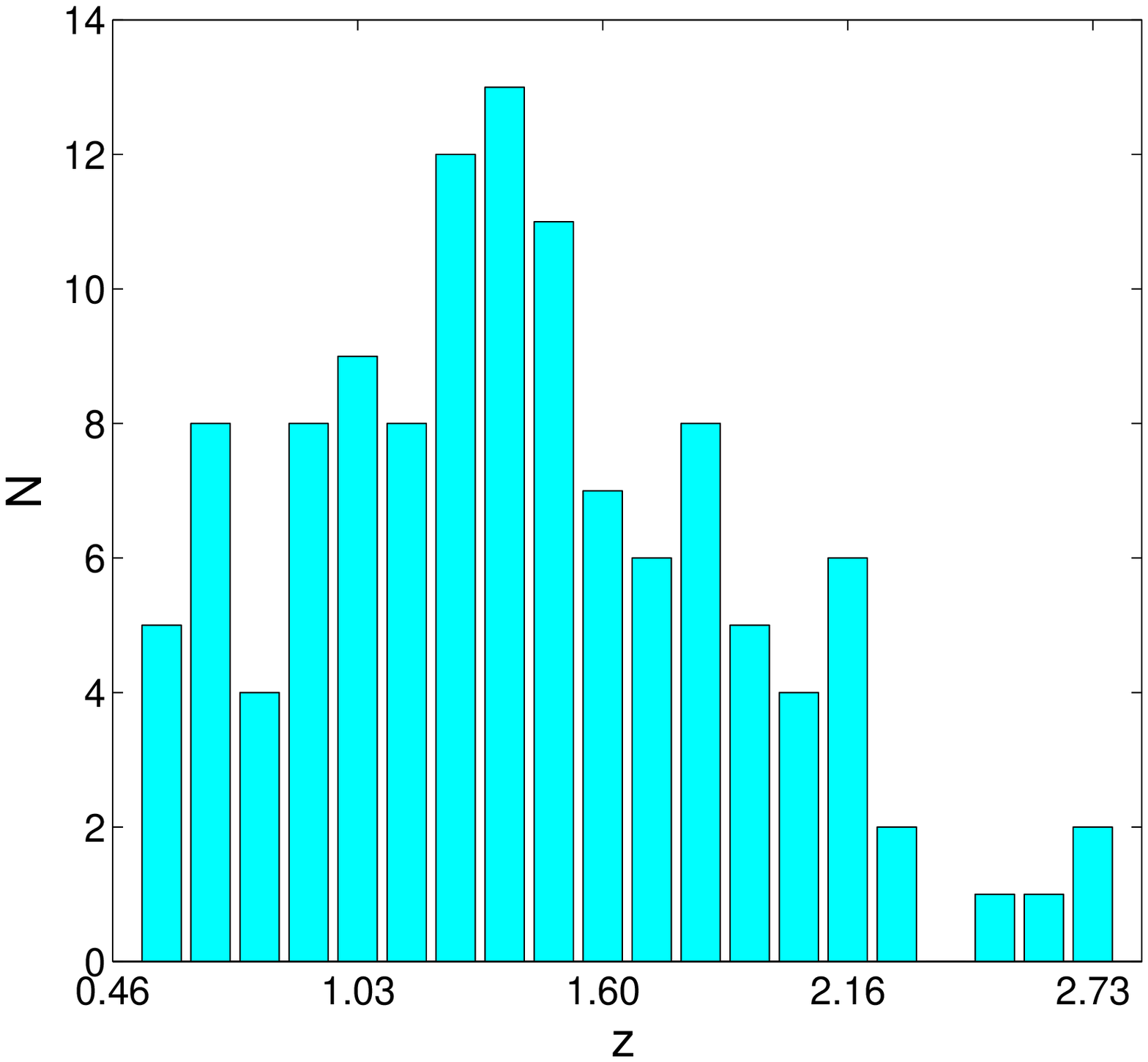} \includegraphics[width=5cm,angle=0]{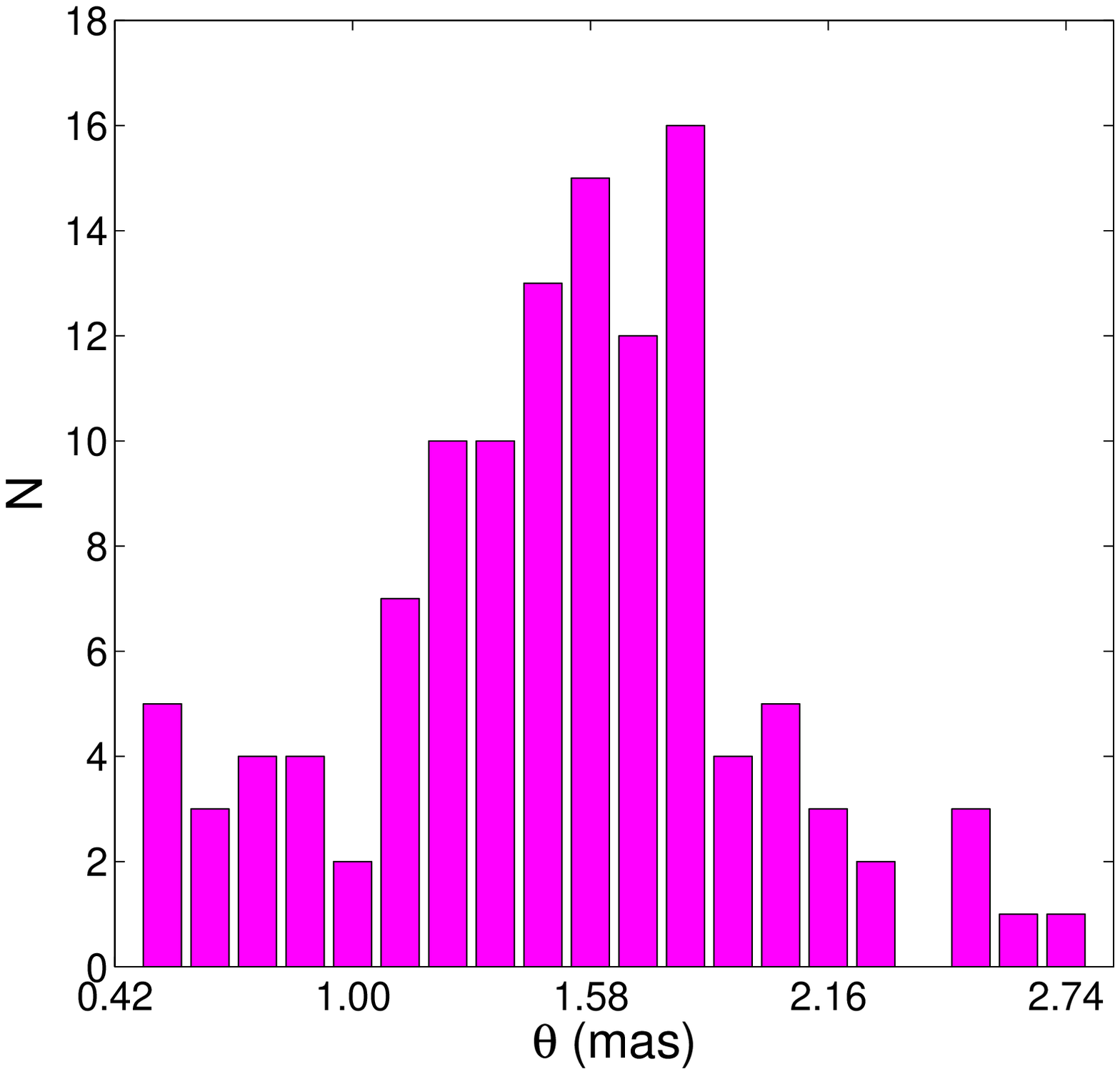} \\
\end{center}
\caption{ Upper: Sky distribution of the quasar sample of 120
sources detected with VLBI. Lower: Luminosity, redshift, and angular
size distribution for the quasar sample of 120
sources.}\label{position}
\end{figure*}

\begin{figure*}
\begin{center}
\includegraphics[width=8cm,angle=0]{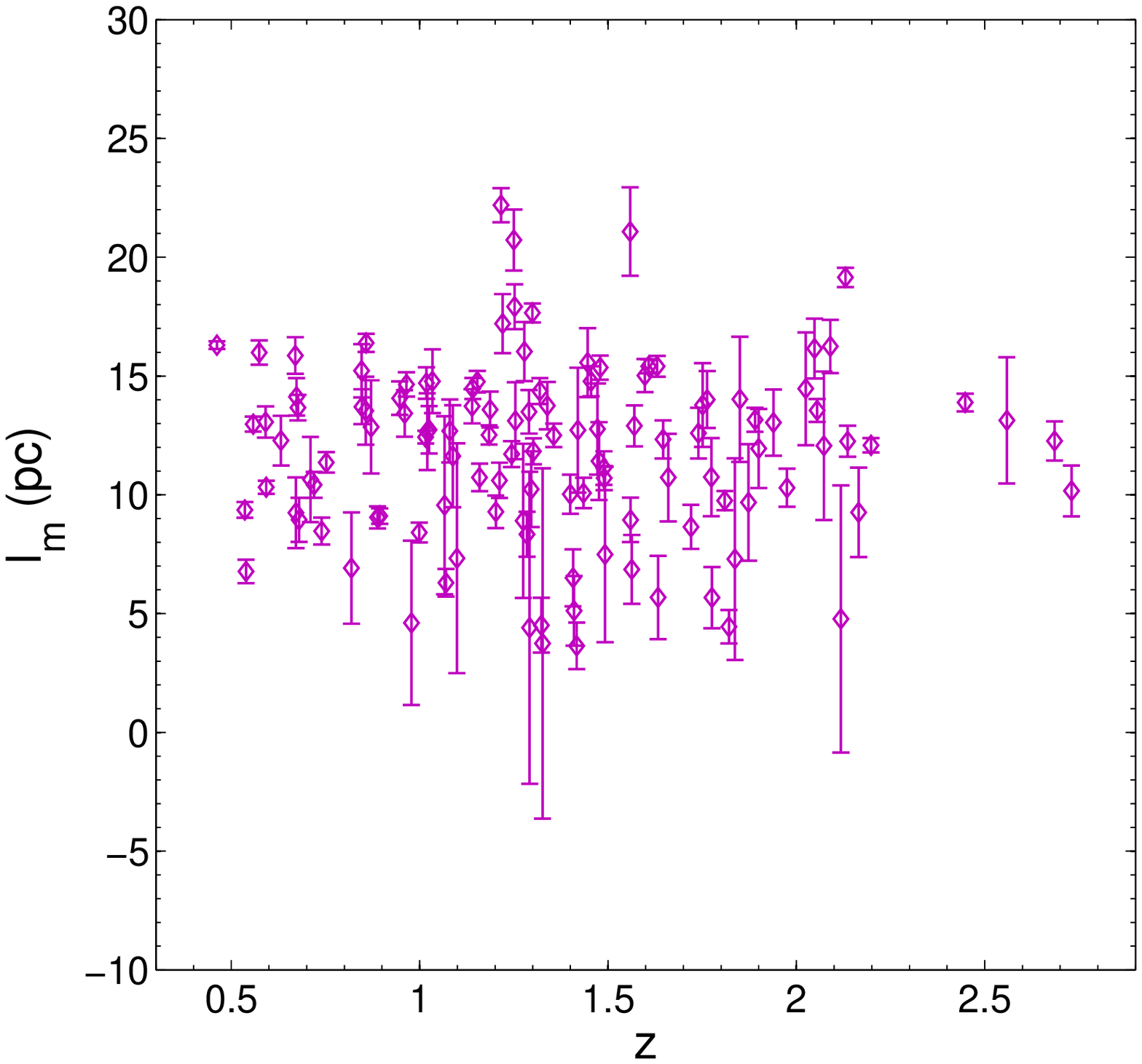} \includegraphics[width=8cm,angle=0]{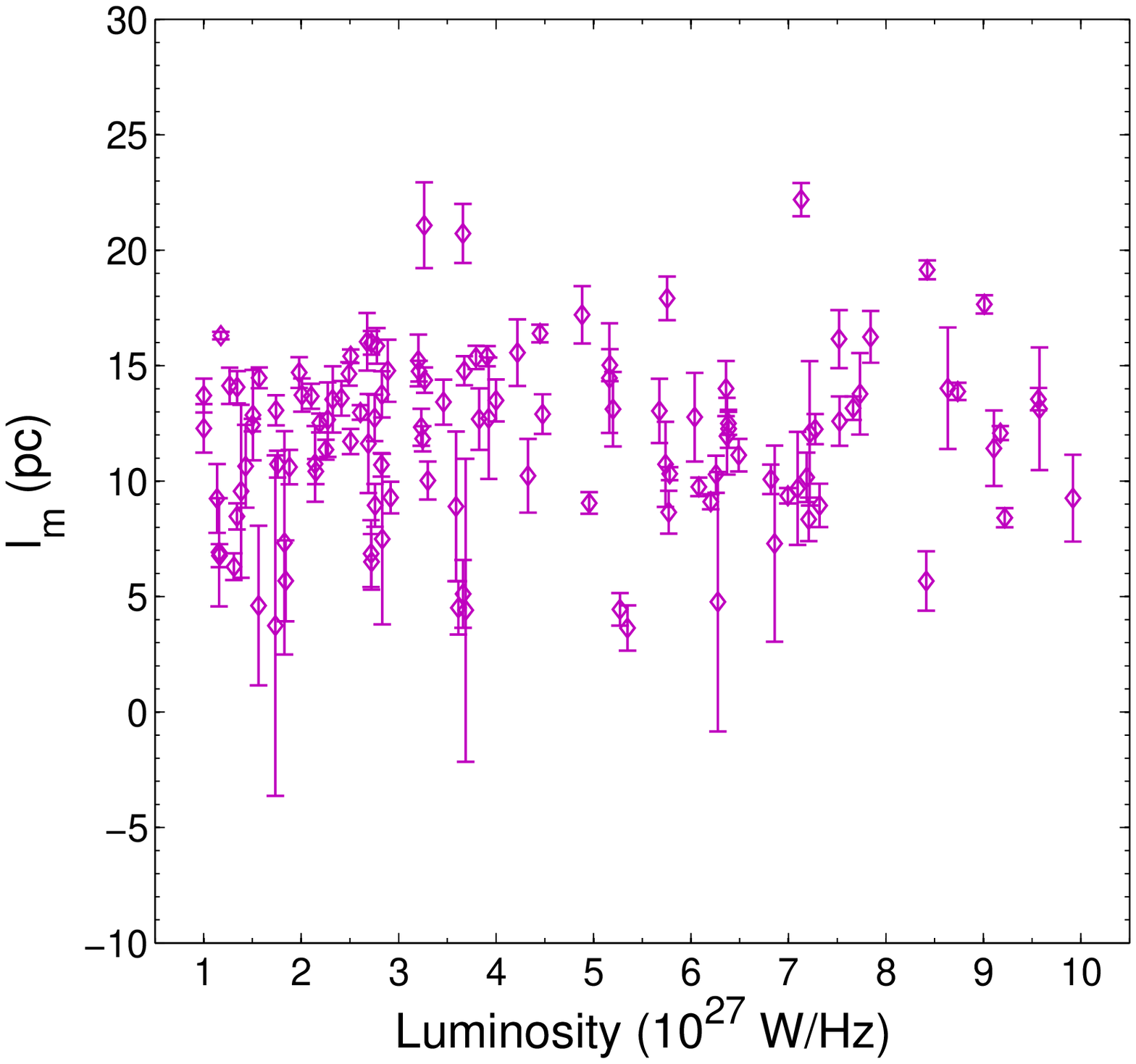}
\end{center}
\caption{ The intrinsic metric linear size $l_m$ distribution of the
120 intermediate luminosity quasars in the redshift and luminosity
space. }\label{ztheta}
\end{figure*}

\begin{figure}
\begin{center}
  \includegraphics[scale=0.45]{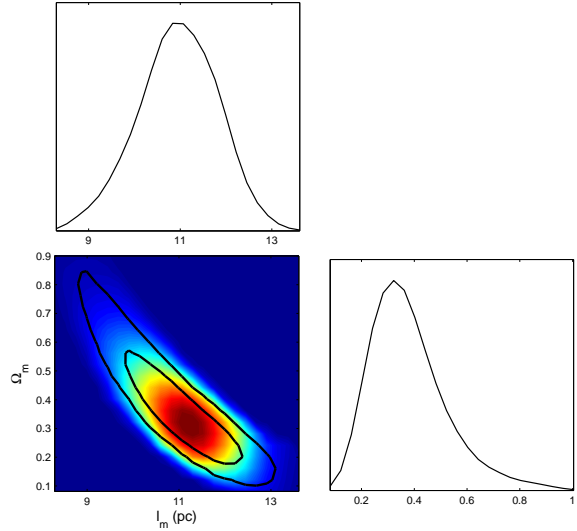}
  \end{center}
  \caption{ Degeneracy between the linear size of ultra-compact structure in radio
  quasars and the matter density parameter in the Universe; $68.3\%$ and $95.4\%$ confidence regions are shown.}\label{lomegam}
\end{figure}

\begin{figure}
\begin{center}
  \includegraphics[scale=0.45]{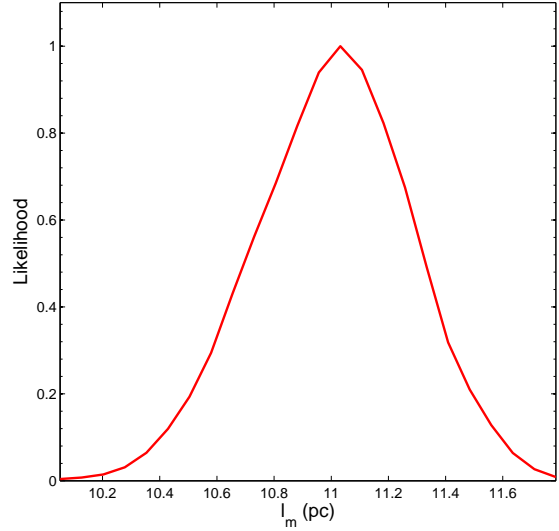}
  \end{center}
  \caption{ The corrected linear size of compact quasars derived in a cosmological model independent way, using the observational $H(z)$ data.
}\label{fig3}
\end{figure}

\begin{figure}
\begin{center}
 \includegraphics[scale=0.45]{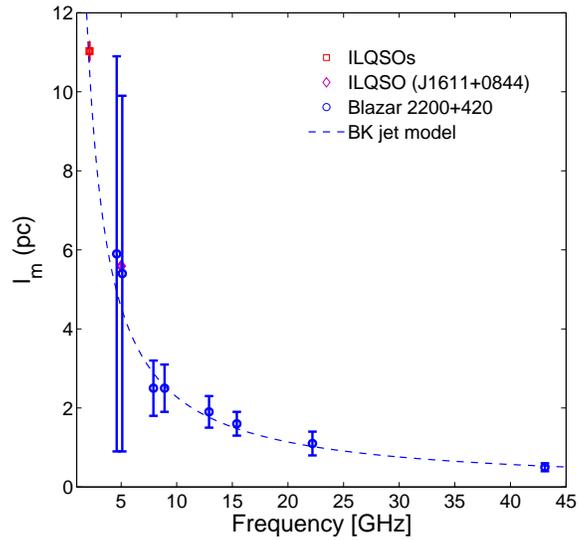}
  \end{center}
  \caption{ The plot of core size versus frequency (Constraint on the linear size of compact structure in intermediate-luminosity quasars $l$ at 2.16 GHz is added).
  BK 79 jet model ($r_c \propto \nu^{-1}$ )is used for the fitted curve from VLBI/VLBA observations in the literature.}\label{BK79}
\end{figure}

\begin{figure*}
\begin{center}
\includegraphics[width=8cm,angle=0]{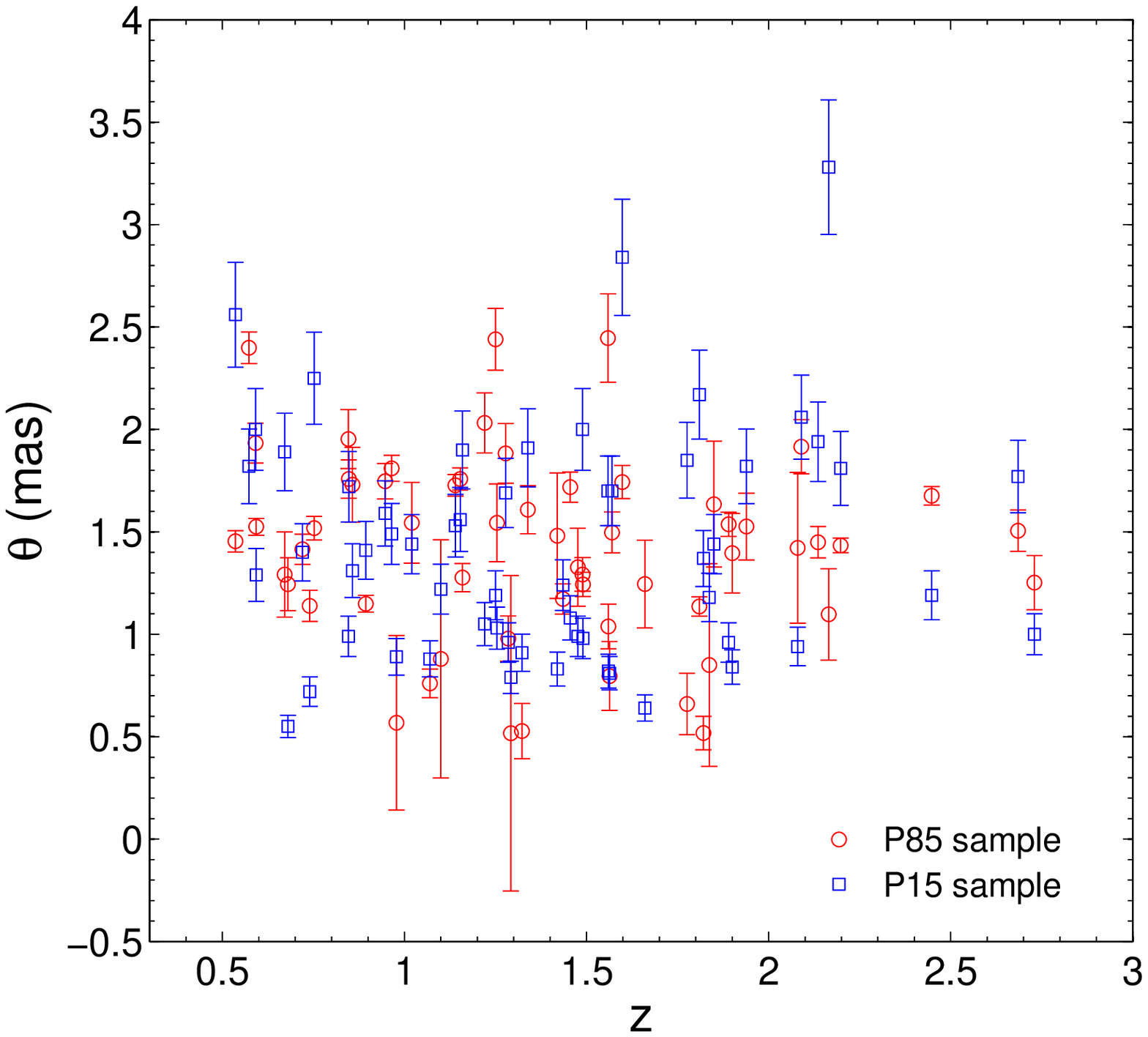}  \includegraphics[width=8cm,angle=0]{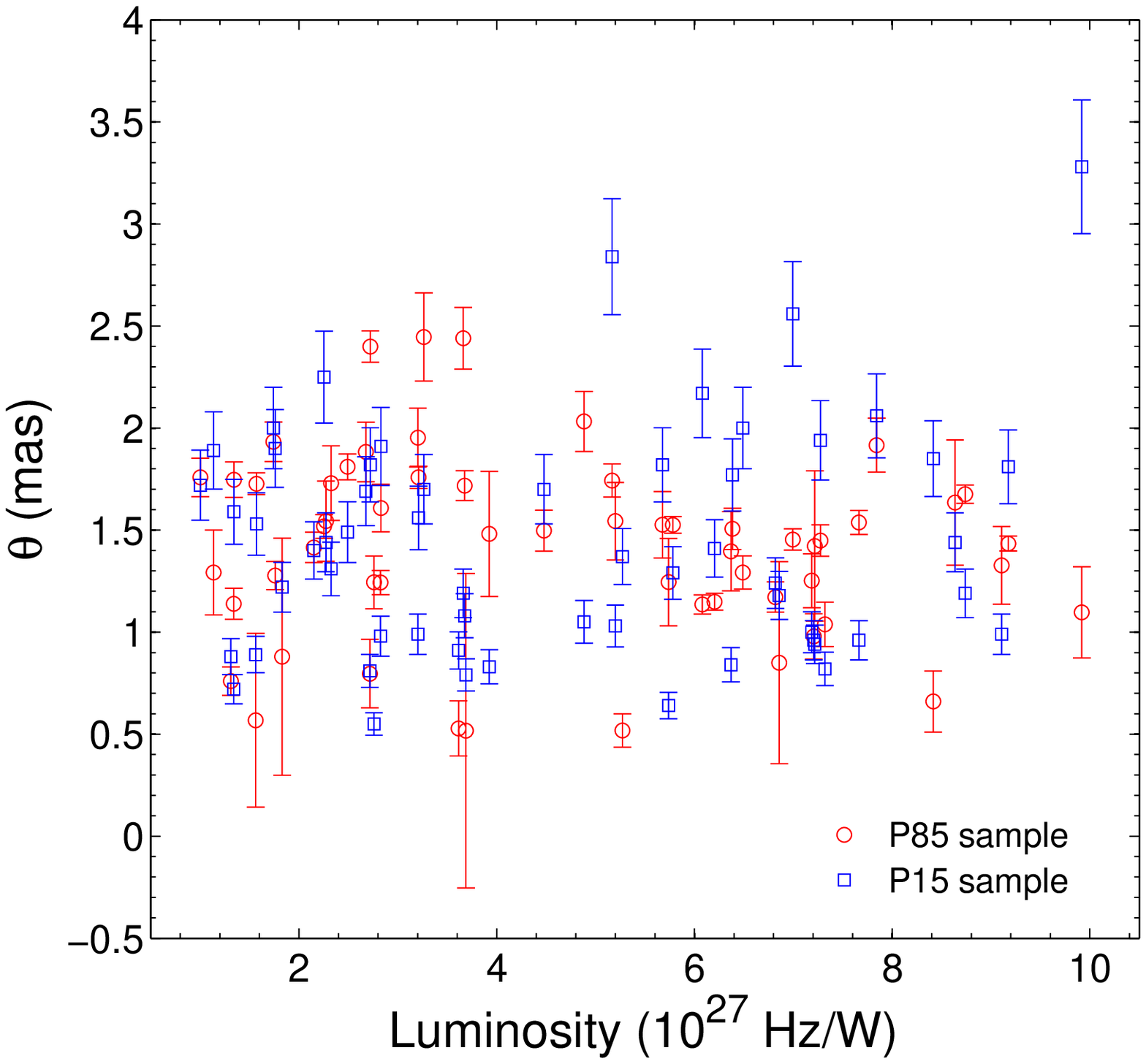} \\
\end{center}
\caption{ Comparison between P15 and P85 samples concerning the
``angular size - redshift'' and ``angular size - luminosity''
diagrams based on 58 sources common to these two surveys.
}\label{P15}
\end{figure*}

\begin{figure}
\begin{center}
  \includegraphics[scale=0.45]{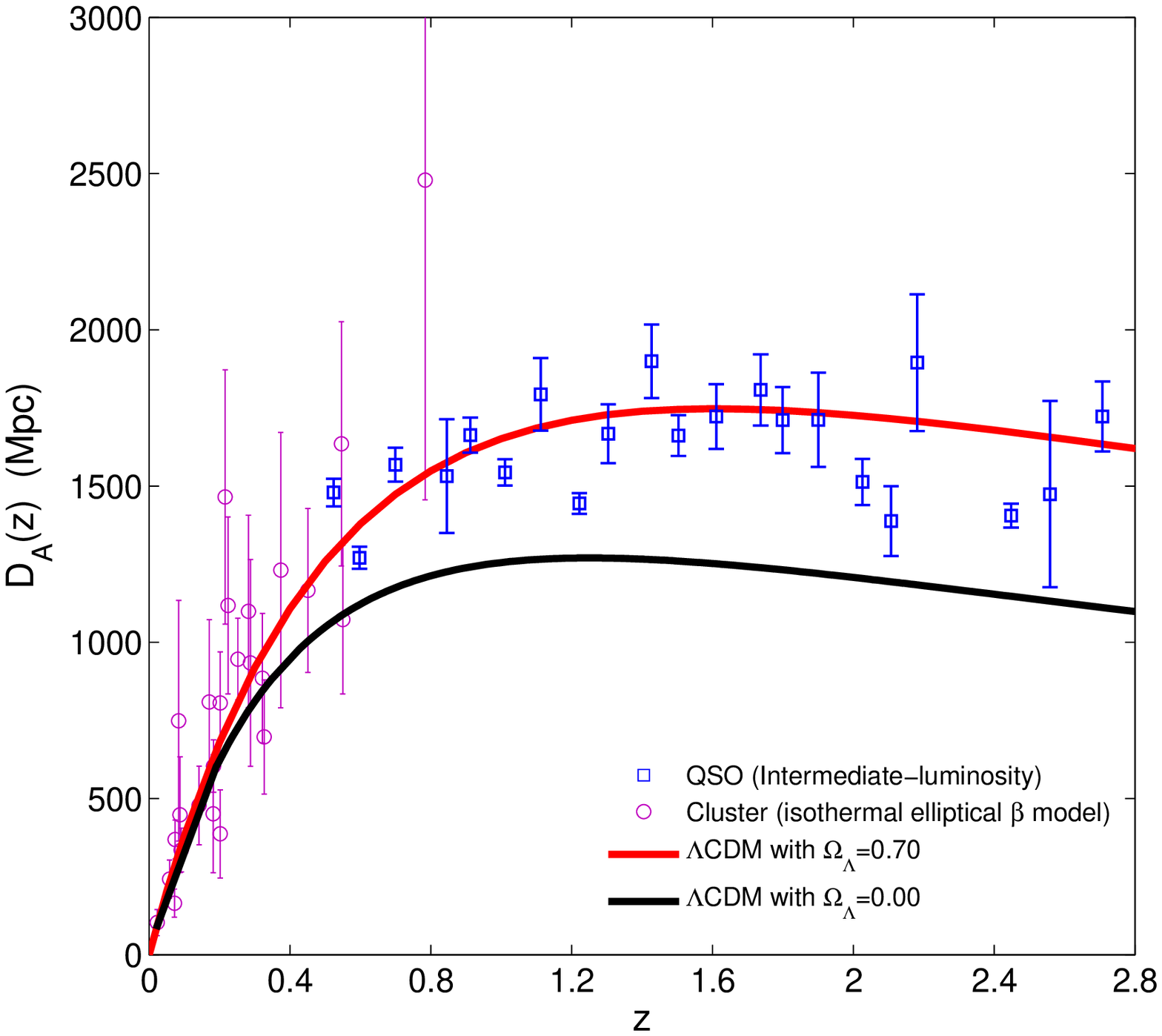}
  \end{center}
  \caption{The angular diameter distances $D_A(z)$ estimated from quasars as standard rulers (blue squares). Angular diameter distances from galaxy clusters (purple circles) are also added for comparison. Theoretical predictions of $\Lambda$CDM models with $\Omega_\Lambda=0.00$ and $\Omega_\Lambda=0.70$ are denoted by black and red solid lines, respectively. }\label{DAcom}
\end{figure}

\section{Data and method} \label{Data}

The data to be used here are derived from a compilation of
angular-size/redshift data for ultra-compact radio sources, from an
ancient VLBI survey undertaken by \citet{Preston85} (hereafter we
call this data P85 for short). By employing a world-wide array of
dishes forming an interferometric system with an effective baseline
of about $8\times 10^7$ wavelengths, this survey succeeded to detect
917 sources with compact structure out of 1398 known radio sources.
The results of this survey were utilized initially to provide a very
accurate VLBI celestial reference frame, improving precision by at
least an order of magnitude, compared with earlier stellar frames.
An additional expectation was that the catalog would be ``used in
statistical studies of radio-source properties and cosmological
models'' \citep{Preston85}. By considering a sample including 258
objects with redshifts $z>0.5$, the possibility of applying this
sample to cosmological study was firstly proposed by
\citet{Gurvits94} and then extended by
\citet{Jackson97,Jackson04,Jackson06,Cao16}. We will use a revised
sample comprising 613 objects with reshifts $0.0035\leq z\leq
3.787$, which sample is a recent upgrade with regard to redshift
based on P85 \citep{Jackson06} \footnote{ A full listing of all 613
objects, with appropriate parameters, is available in electronic
form via http://nrl.northumbria.ac.uk/13109/.}.

All detected sources included in this comprehensive compilation were
imaged with VLBI at 2.29 GHz, which involve a wide class of
extragalactic objects including quasars, radio galaxies, BL Lac
objects (blazars), etc. It should be noted that P85 does not give
contour maps, and does not list angular sizes explicitly; however,
total flux density and correlated flux density (fringe amplitude)
are listed; the ratio of these two quantities is the visibility
modulus $\Gamma$, which defines a characteristic angular size
\begin{equation}
\theta={2\sqrt{-\ln\Gamma \ln 2} \over \pi B} \label{thetaG}
\end{equation}
where $B$ is the interferometer baseline, measured in wavelengths
\citep{Thompson86,Gurvits94}.  The angular sizes used here were
calculated using equation (\ref{thetaG}); it is argued in
\citet{Jackson04} that this size represents that of the core, rather
than the angular distance between the latter and a distant weak
component.

In our analysis, by applying two selection criteria, one on spectral
index ($-0.38\leq \alpha\leq 0.18$) and the second on luminosity
($10^{27}$ W/Hz$<L<$$10^{28}$ W/Hz), we will focus our attention on
the compact structure in 120 radio quasars with flat spectral index
and intermediate luminosity. Recent study of \citet{Cao16} suggested
that they can be effectively used as standard rulers. Full
information about the said 120 sources can be found in Table 1,
including source coordinates, redshifts, angular size, spectral
index, and total flux density. The corresponding optical counterpart
for each system can be found in P85. Fig.~\ref{position} is an
equal-area sky-distribution plot of the detected quasars. The
distribution of redshifts, luminosities, and angular-sizes of the
sources in our sample is also shown in Fig.~\ref{position}, which
simply reflects the fact that our basically luminosity-limited
sample, compiled on an ad-hoc basis from the literature and based
upon various selection criteria, is relatively homogenous both in
redshifts and angular-sizes. 
The angular sizes of the sample range from 0.424 to 2.743
milliarcsec, with 15\% of the quasars having angular sizes
$\theta<1$ mas, and only a handful of quasars with larger angular
sizes ($\theta>2$ mas) have been identified, while 75\% of all
quasars are located at 1.0 mas$<\theta<2.0$ mas. We remark here
that, the final sample covers the redshift range $0.46<z<2.76$,
which indicates its potential usefulness in cosmology at high
redshifts.

For a cosmological rod with intrinsic length $l_m$, the angular
size-redshift relation can be written as \citep{Sandage88}
\begin{equation}
\theta(z)= \frac{l_m}{D_A(z)} \label{theta}
\end{equation}
where $\theta(z)$ is the angular size at redshift $z$, and $D_A(z)$
is the corresponding angular diameter distance. $D_A(z)$ is related
to $H_0$, the Hubble constant, and $E(z; \textbf{p})$, the
dimensionless expansion rate depending on redshift $z$ and
cosmological parameters $\textbf{p}$. However, the cosmological
application of such technique requires good knowledge of the linear
size of the ``standard rod'' used. The possibility that source's
linear size depends on the source luminosity and redshift should be
kept in mind. In this analysis, we use a phenomenological model to
characterize the relations between the projected linear size of a
source and its luminosity $L$ and redshift $z$
\citep{Gurvits94,Gurvits99,Cao16}
\begin{equation} \label{lm}
l_m=lL^\beta(1+z)^n\,
\end{equation}
where $l$ is the linear size scaling factor, $\beta$ and $n$
quantify the dependence of the linear size on source luminosity and
redshift, respectively \citep{Gurvits94}. Following the analysis of
\citet{Cao16}, for our quasar sample, the linear size $l_m$ is
independent of both redshift and luminosity ($|n|\simeq 10^{-3}$,
$|\beta|\simeq 10^{-4}$) and there is only one parameter $l$ to be
considered. The intrinsic metric linear size $l_m$ distribution of
the 120 intermediate luminosity quasars in the redshift and
luminosity space is shown in Fig.~2. Since both the $l_m$
(calculated from Eq.~\ref{theta}) and luminosity require the
knowledge of angular diameter distances, we used $D_A(z)$ values at
the quasars' redshifts inferred from the Hubble parameter ($H(z)$)
measurements, i.e. in a cosmological model independent way (see
\citet{Cao16} for details). The measurements of $H(z)$ are acquired
by means of two different techniques: one is called cosmic
chronometers~\citep{JimenezLoeb}, i.e. massive, early-type galaxies
evolving passively on a timescale longer than their age difference,
while the other comes from the analysis of baryon acoustic
oscillations (BAO). One can see from Fig.~2 that
intermediate-luminosity quasars could serve as standard rulers, much
better than other quasar sub-samples discussed in \citet{Cao16}.
Therefore, if we could find a suitable method to calibrate $l$, then
we would get stringent constraints on the angular diameter distances
at different redshifs, and thus relevant cosmological parameters
$\textbf{p}$. In order to check the degeneracy between $\textbf{p}$
and $l$, we investigated our radio quasar sample in the framework of
the concordance $\Lambda$CDM cosmology, which is characterized by
two free parameters, the matter density $\Omega_m$ and the intrinsic
linear size $l$. Fig.~\ref{lomegam} shows the corresponding
confidence regions.

Now a cosmological-model-independent method will be applied to
derive the linear size of the compact structure in
intermediate-luminosity radio quasars. Namely, cosmic chronometer
$H(z)$ measurements processed using Gaussian Processes (GP)
\citep{Li15} can provide us angular diameter distances $D_A$
covering the quasar redshift range, and thus allows us to calibrate
the angular size of milliarcsecond quasars. In the so-called cosmic
chronometers~\citep{JimenezLoeb}, the cosmic expansion rates $H(z)$
are measured from age estimates of red galaxies without any prior
assumption of cosmology, i.e., $H(z)\simeq-\frac{1}{1+z}\frac{\Delta
z}{\Delta t}$. In order to minimize the systematic effects, we used
the $H(z)$ sample following the choice of
\citet{morescoetal,licia2014,Li15}. Currently, based on this method,
30 measurements of $H(z)$ covering the redshift range $0.070 \leq
z\leq 1.965$ have been obtained. See \citet{Zheng2016} for details
and reference to the source papers. However, according to the
analysis of \citet{morescoetal}, the choice of stellar population
synthesis model may strongly affect these estimates of $\Delta t$
and thus $H(z)$, especially at $z \geq 1.2$. Therefore, we consider
only 24 $H(z)$ measurements up to $z < 1.2$ in this paper. Moreover,
following the analysis of \citet{licia2014,Li13}, the error bar of
the highest-$z$ point is increased by 20\% to include the
uncertainties of the stellar population synthesis models.

According to \citet{Holanda13}, for a non-uniformly distributed
$H(z)$ data, the comoving distance integral could be obtained with a
simple trapezoidal rule
\begin{equation}\label{integral}
D(z)=c\int_0^z\frac{dz'}{H(z')}\approx\frac{c}{2}\sum_{i=1}^N(z_{i+1}-z_i)
\bigg(\frac{1}{H_{i+1}}+\frac{1}{H_i}\bigg).
\end{equation}
Concerning the fact that the number of data points and the
uniformity of the spaced data will heavily influence the precision
of this simple rule, we will use Gaussian Processes (GP), a powerful
non-linear interpolating tool to reconstruct the evolution of the
expansion rate with redshift, and thus integrate its inverse
function to estimate distances in a cosmological model-independent
way \citep{Holanda13}. This method was firstly proposed to test both
cosmology \citep{Holsclaw10a,Holsclaw10b} and
cosmography~\citep{Shafieloo13}, and then extensively applied to the
derivation of the Hubble constant $H_0$ \citep{Busti14}, the
reconstructions of the equation of state of dark
energy~\citep{Seikel12a} and the distance-duality
relation~\citep{Zhang14}. The advantage of Gaussian processes is,
that we do not need to assume any parametrized model for $H(z)$
while reconstructing this function from the data
\citep{Holsclaw10a,Holsclaw10b}. Moreover, with a very small and
uniform step of $\Delta z=z_{i+1}-z_i$,  we may obtain more precise
measurements of angular diameter distances at a certain redshift. In
order to reconstruct the Hubble parameter as a function of the
redshift from 24 $H(z)$ measurements of cosmic chronometers covering
$z=(0.0,1.2)$ we used the publicly available code \citep{Seikel12a}
called the GaPP (Gaussian Processes in Python)
\footnote{http://www.acgc.uct.ac.za/$\sim$seikel/GAPP/index.html}.

Applying the redshift-selection criteria, $\Delta
z=|z_{QSO}-z_{H}|\leq 0.005$ to the angular diameter distances
derived from $H(z)$, we obtain 48 measurements of $D_A$ coinciding
with the quasars reshifts. Undertaking similar analysis as
\cite{Cao16}, we obtain constraints on the linear size $l_m$ with
the best fit
\begin{eqnarray}
&& l_m = 11.03\pm0.25 \ \mathrm{pc}. 
\end{eqnarray}
The probability distribution of $l_m$ is also shown in
Fig.~\ref{fig3}, which will be used in the following cosmological
analysis with quasar observations. Note, that in our previous paper
\citep{Cao16} which aimed at estimation of the velocity of light
with extragalactic sources, such calibration was not be possible
because of the appearance of $c$ in the expression for $D_A$ (see
e.g. Eq.~\ref{integral}). Two issues deserve attention. First is
that the rigid assumption of vanishing $\beta$ and $n$ parameters
describing evolution of the comoving size $l_m$ with luminosity and
redshift might introduce a bias and underestimate the uncertainties
of cosmological parameters fitted. Second is whether the cosmic
chronometers used for calibrating QSO as standard rulers introduce a
bias. These points will be addressed in the following sections by
modelling $\beta$ and $n$ parameters with Gaussian distributions and
by comparison of the results obtained with quasars as standard
rulers and with $H(z)$ data alone.

Now we try to make some comments on the physical meaning of the
linear size of this standard ruler. For a long time it has been
argued that Active Galactic Nuclei (AGN) must be powered by
accretion of mass onto massive black holes. Current theoretical
models indicate that jets of relativistic plasma are generated in
the central regions of AGN, and magnetic fields surrounding the
black hole expel, accelerate and help to collimate the jet flow
outwards \citep{Meier09}. According to the unified classification of
Active Galactic Nuclei (AGN), 10 pc is the typical radius at which
AGN jets are apparently generated and there is almost no stellar
contribution \citep{Blandford78}. In the conical jet model proposed
by \citet{Blandford79} [hereafter BK79], with the base of the jet
corresponding to the vertex of the cone, the unresolved core is
identified with the innermost, optically thick region of the
approaching jet. For QSOs, this compact opaque parsec-scale core is
located between the broad-line region ($\sim$ 1 pc) and narrow-line
region ($\sim$ 100 pc) \citep{Blandford78}.

More recently, \citet{Hopkins10} have investigated the correlation
between the black hole's mass accretion rate $\dot{M}_{BH}$ and the
star-formation rate $\dot{M}_\ast$. Their results from simulations
showed that, within the central 10 pc around the black hole, the
star formation rate is equal to the mass accretion rate. This
conclusion is well-consistent with the recent observations by
\citet{Silverman09}. \citet{Frey10} has presented high-resolution
radio structure imaging of five quasars (J0813+3508, J1146+4037,
J1242+5422, J1611+0844, and J1659+2101) at $4.5<z<5$ with the
European VLBI Network (EVN) at 5 GHz. Although all cases have
satisfied the intermediate-luminosity criterion defined in this
paper, there is only one flat-spectrum source (J1611+0844) which
could be used for comparison in our analysis, while the compact
emission of other quasars is characterized by a steep radio spectrum
\citep{Frey10}. Moreover, we combine our constraint with the recent
observation of Blazar 2200+420 from the Very Long Baseline Array
(VLBA), at eight frequencies (4.6, 5.1, 7.9, 8.9, 12.9, 15.4, 22.2,
43.1 GHz), to investigate the frequency-dependent position of VLBI
cores \footnote{We use the term ``core'' as the apparent origin of
AGN jets that commonly appears as the brightest feature in VLBI
images of blazars \citep{Lobanov98}.}. From the comparison presented
in Fig.~\ref{BK79} and Table 2, we find a very good consistency
between the measurements and the BK79 conical jet model, in which
the position of the radio core follows $r \propto \nu^{-1}$, where
$r$ is the distance from the central engine \citep{Blandford79}, if
the core is self-absorbed and in equipartition. As noted in the
analysis of \citet{Cao16}, our estimate of $l_m$ is also well
consistent with the results derived from recent multi-frequency VLBI
imaging observations of more than 3000 compact extragalactic radio
sources \citep{Pushkarev15}. In fact, 58 intermediate-luminosity
quasars included in our modified P85 sub-sample have also been
observed by recent VLBI observations based on better uv-coverage in
the P15 sample. Based on these 58 sources, one may compare P15 and
P85 samples on the ``angular size - redshift'' and ``angular size -
luminosity'' diagrams. This is displayed in Fig.~\ref{P15}, from
which one can see some difference between the samples concerning
estimates of the angular size. However, we checked that the
characteristic linear size $l_m$ at 2 GHz estimated from P15 is
well-consistent with the results obtained from the P85 sample.
Astrophysical application of the recent multi-frequency angular size
measurements of 58 intermediate-luminosity quasars from the P15
sample, will be subject of the next work in preparation
\citep{Cao17b}.


If one, quite straightforwardly, attempts to construct an empirical
relation $D_A(z)$ extending to higher redshifts on the basis of
individual quasar angular sizes, using Eq.~(\ref{theta}), one can
obtain the $D_A$ measurement and the corresponding uncertainty for
each quasar. However, this procedure results in large uncertainties
in $D_A$, which problem has been encountered previously
\citep{Gurvits94,Gurvits99}, as can be seen from plots of the
measured angular size against redshift therein. This problem remains
even after 13 systems with very large ($\sim 50\%$) uncertainties
are removed. Therefore, in order to minimize its influence on our
analysis, we have chosen to bin the remaining 107 data points and to
examine the change in $D_A$ with redshift. The final sample was
grouped into 20 redshift bins of width $\Delta z=0.10$.
Fig.~\ref{DAcom} shows the median values of $D_A$ for each bin
plotted against the central redshift of the bin. For comparison, the
two curves plotted as solid lines represent theoretical expectations
from the concordance $\Lambda$CDM model and the Einstein - de Sitter
model. One can see that the latter is disfavored at high confidence.
More importantly, the angular diameter distance information obtained
from quasars has helped us to bridge the ``redshift desert'' and
extend our investigation of dark energy to much higher redshifts. It
is worth noting that these 120 intermediate-luminosity QSOs are
obtained in a completely cosmology-independent method, and hence can
be used to constrain cosmological parameters without the circularity
problems.

\section{Constraints on dark energy} \label{Cosmology}

In this section, we investigate some dark energy models and estimate
their best-fitted parameters using the quasar sample. Models
discussed in this section are aimed at explaining the accelerated
expansion of the universe by introducing a hypothetical fluid whose
contribution to the matter budget and equation of state are unknown
parameters to be fitted. Next section~\ref{Alternative}, will
discuss alternative concepts involving departure from classical
General Relativity. More specifically, the following models for dark
energy will be studied in this section:
\begin{itemize}
\item[] $\Lambda$CDM:  cosmological constant in a flat universe.
\item[] XCDM:  constant equation-of-state parameter in a flat universe.
\item[] $w_z$CDM:  time-varying equation-of-state parameter in a flat universe.
\end{itemize}
We determine the cosmological model parameters \textbf{p} using a
$\chi^{2}$ minimization method.
\begin{equation}
\label{eq:chi2} \chi^{2}(\textbf{p}) =
  \sum_{i}^{120}{\frac{\left[\theta(z_{i}; l_m; \textbf{p})
     - \theta_{oi}\right]^{2}}{\sigma_{i}^{2}}},
\end{equation}
where $\theta(z_{i}; \textbf{p}) = l_m/D_A$ is the angle subtended
by an object of proper length $l_m$ transverse to the line of sight
and $\theta_{oi}$ is the observed value of the angular size with
uncertainties $\sigma_{i}$. The summation is over all the 120
observational data points. In computing $\chi^2$ we have also
assumed additional 10\% uncertainties in the observed angular sizes,
to account for both observational errors and the intrinsic spread in
linear sizes. We remark here that, although the best-fit values of
$\beta$ and $n$ parameters, describing the dependence of $l_m$ on
the luminosity and redshift, are negligibly small, yet their
uncertainties could also be an important source of systematic errors
on the final cosmological results. In order to address this issue,
we perform a sensitivity analysis by applying Monte Carlo
simulations in which $\beta$, $n$ were characterized by Gaussian
distributions: $\beta=0.00\pm0.05$ and $n=0.00\pm0.05$, while the
uncertainty of the linear size scaling factor was taken into account
with a Gaussian distribution as $l=11.03\pm0.25$ pc \footnote{Note
that the additional 10\% uncertainties in the observed angular sizes
applied to $\chi^{2}$ minimization method is equivalent to adding an
additional 10\% uncertainty in the linear size scaling factor.}.

In a similar manner as in other papers introducing new compilations
of cosmologically important data sets, e.g. \citet{Amanullah10}, we
constrain the properties of dark energy first using QSO alone (with
and without the systematic uncertainty of $l$, $\beta$ and $n$), and
then perform combined analysis using also the latest CMB data from
\citet{Ade16}, and the BAO data from 6dFGS, SDSS-MGS, BOSS-LOWZ, and
BOSS-CMASS \citep{Beutler11,Ross15,Anderson14}. Moreover,
considering that there is no strong evidence for the departure from
spatially flat geometry at the current data level, which is known
from and strongly supported by other independent and precise
experiments \citep{Ade16a,Ade16}, we will assume spatial flatness of
the Universe throughout the following analysis in the paper. The
results for each of the models are listed in Table~2 and discussed
in turn in the following sub-sections. Unless stated otherwise, the
uncertainties represent the 68.3\% confidence limits and include
both statistical uncertainties and systematic errors. In addition,
we add the prior for the Hubble constant  $H_0=67.3 \;
\rm{kms}^{-1}\rm{Mpc}^{-1}$ after Planck Collaboration XVI (2014).
The exceptions are the $\Lambda$CDM and the DGP models (next
section) where $H_0$ is treated as a free parameter.

\subsection{$\Lambda$CDM}

\begin{figure*}
\begin{center}
  \includegraphics[width=8cm,angle=0]{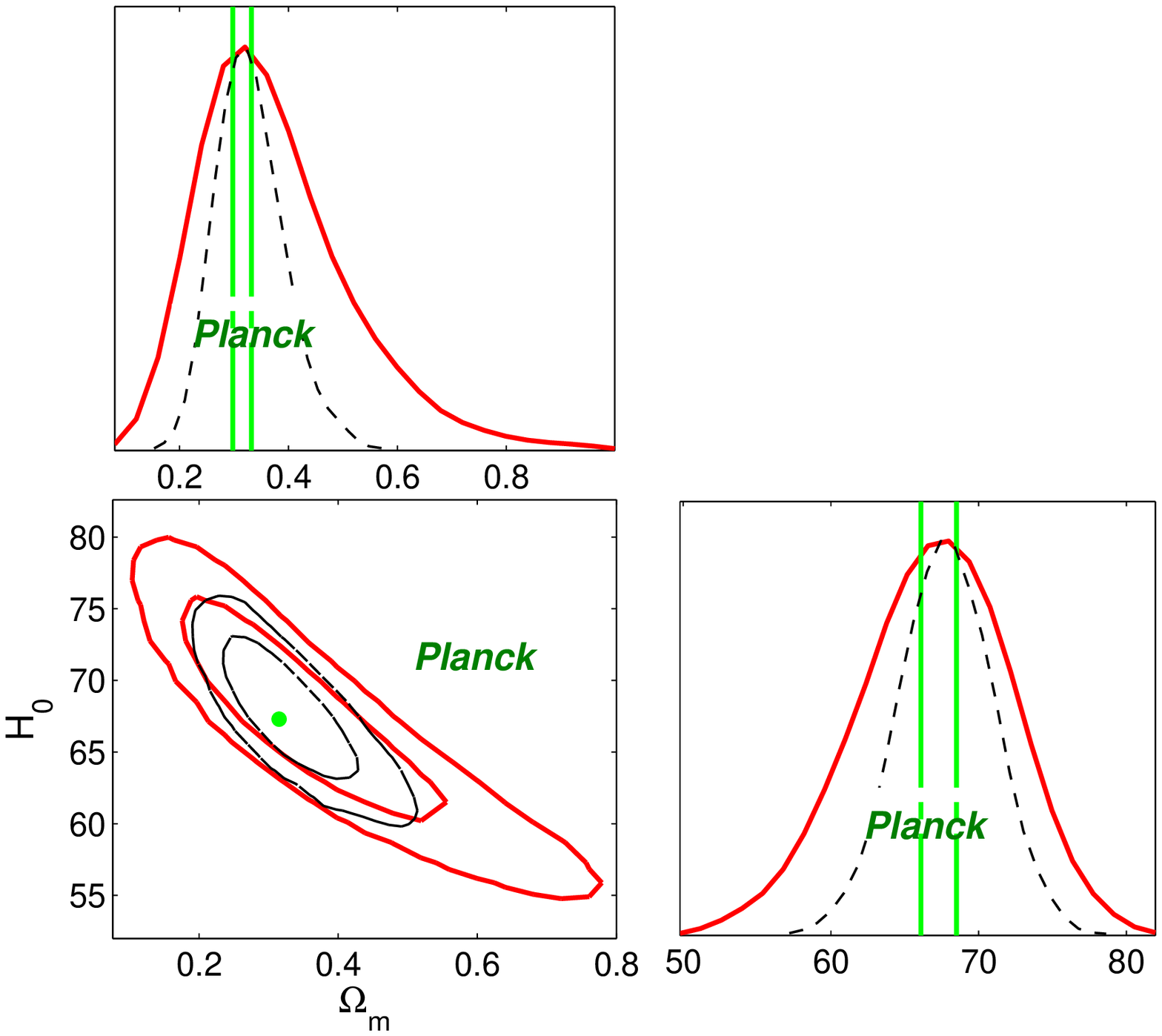} \includegraphics[width=8cm,angle=0]{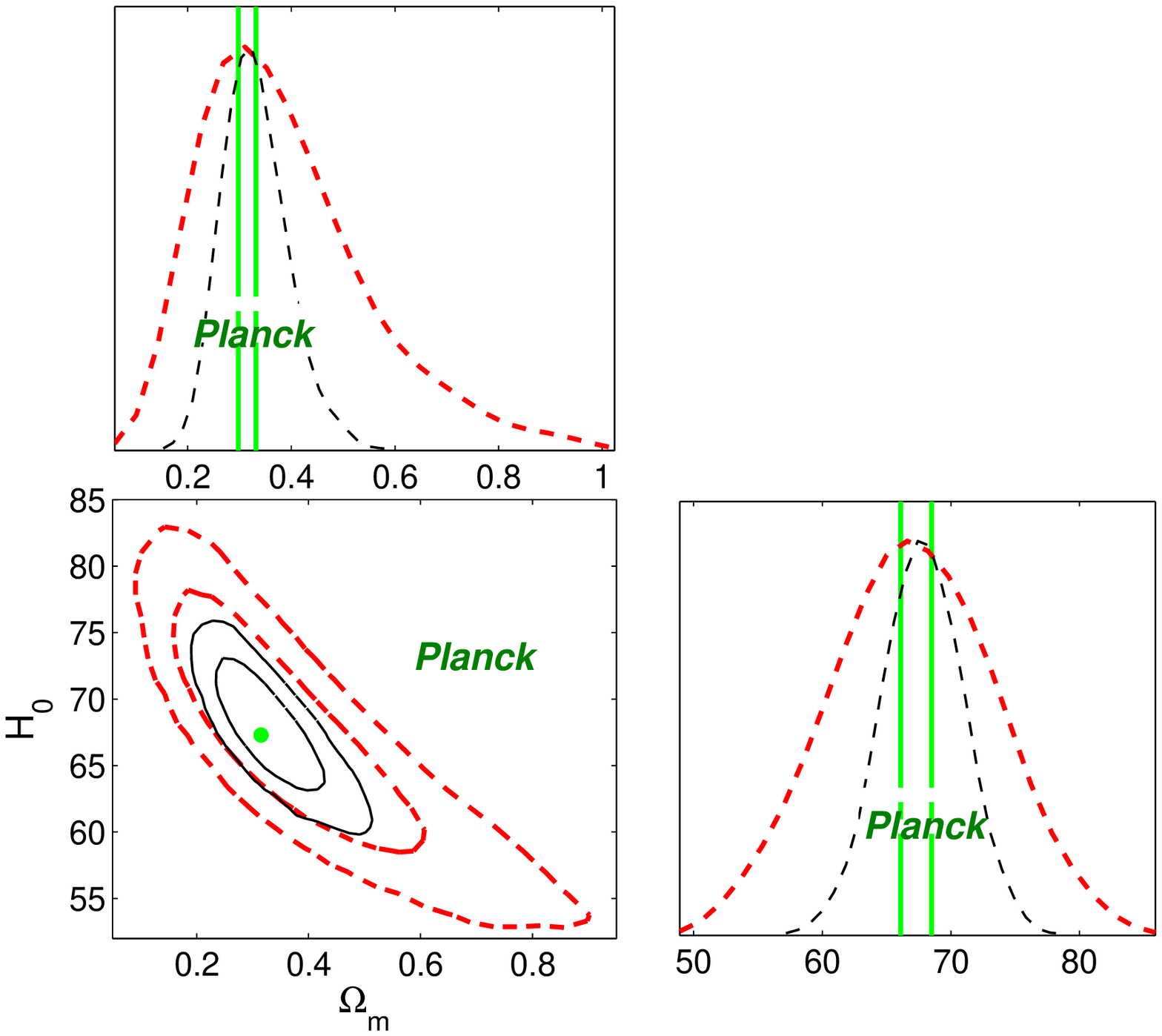}
  \end{center}
  \caption{ Cosmological constraints on the flat $\Lambda$CDM model from the quasar sample without (left
panel) and with systematical uncertainties (right panel). Fitting
results from recent $H(z)$ measurements (black dashed lines) and
Planck observations (green dot represents the best-fist with
$1\sigma$ errors denoted by green solid lines) are also added for
comparison. }\label{LCDM}
\end{figure*}

\begin{figure*}
\begin{center}
  \includegraphics[width=7cm,angle=0]{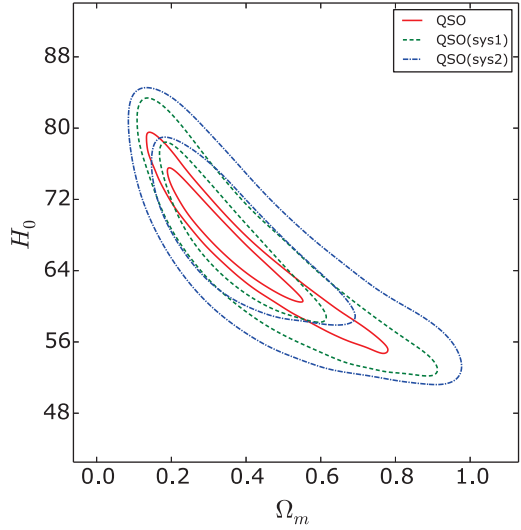}
  \end{center}
\caption{ Cosmological constraints on the flat $\Lambda$CDM model
from the quasar sample without and with different systematical
uncertainties: sys1 denotes the systematics with $l=11.03\pm0.25$pc,
$\beta=0.00\pm0.05$ and $n=0.00\pm0.05$, while sys2 represents the
systematics with $l=11.03\pm0.25$pc, $\beta=0.00\pm0.10$ and
$n=0.00\pm0.10$.  }\label{LCDM3}
\end{figure*}

If flatness of the FRW metric is assumed, the only cosmological
parameter of this model is ${\bf{p}}=\left\{\Omega_m\right\}$.
However, in order to check the constraining power of our quasar data
on the Hubble constant, we choose to take $H_0$ as the other free
parameter and obtained $\Omega_m=0.322^{+0.244}_{-0.141}$ and
$H_0=67.6^{+7.8}_{-7.4} \; \rm{kms}^{-1}\rm{Mpc}^{-1}$. After
including systematics due to uncertainties on $l$, $\beta$ and $n$,
the matter density parameter and Hubble constant respectively change
to $\Omega_m = 0.312^{+0.295}_{-0.154}$ and $H_0=67.0^{+11.2}_{-8.6}
\; \rm{kms}^{-1}\rm{Mpc}^{-1}$. These results are presented in
Fig.~\ref{LCDM}. Now one issue which should be discussed is how much
the cosmological parameters are affected by larger uncertainties of
$\beta$ and $n$. For this purpose, we changed the uncertainty of the
luminosity-dependence parameter to $\beta=0.00\pm0.10$ and the
redshift-dependence parameter to $n=0.00\pm0.10$. The comparison of
the resulting constraints on $\Omega_m$ and $H_0$ based on different
systematical uncertainties is shown in Fig.~\ref{LCDM3}. One can
easily check that reduction of the error of $\beta$ and $n$ will
lead to more stringent cosmological fits, which motivates us to
improve constraints on the two parameters with a larger quasar
sample from future VLBI observations based on better uv-coverage
\citep{Pushkarev15}.

Another important issue is the comparison of our cosmological
results with those of earlier studies done using other, alternative
probes. We start by comparing our results with fits obtained using
$H(z)$ measurements from cosmic chronometers. Respective likelihood
contours obtained with the latest $H(z)$ data comprising 30 data
points \citep{Zheng2016} are also plotted in Fig.~\ref{LCDM}. We see
that 1$\sigma$ confidence regions from these two techniques overlap
very well with each other. This means that the results obtained on
the sample of quasars are well consistent with the $H(z)$ fits,
although with larger uncertainties due to systematic uncertainties
of the parameters characterizing this standard ruler. Central fits
however, are almost the same. Besides, looking at the constraints
obtained with QSO and $H(z)$ data, we find similar degeneracy
between $\Omega_m$ and $H_0$. Moreover, the constraint on $\Omega_m$
derived from the mean observed separation of the radio lobes
\citep{Daly03,Guerra98,Guerra00} is in broad agreement with the
results we report. Then, gravitational lensing systems with QSO
acting as sources may provide us another probe of angular diameter
distance data in cosmology, since strong gravitational lensing
statistics depends on the angular diameter distances between the
source, the lens, and the observer. Using the redshift distribution
of radio sources, \citet{Chiba99} calculated the absolute lensing
probability for both optical and radio lenses. The best-fit mass
density obtained in their analysis in a flat cosmology,
$\Omega_m=0.3^{+0.2}_{-0.1}$, is consistent with our results.
Constraints on cosmological models using strong lensing statistics
have been obtained e.g. in \citet{Biesiada10,Cao12b,Cao12c,Cao15b}.
At last, based on the first-year \textit{Planck} results, Planck
Collaboration XVI (2014) gave the best-fit parameter:
${\Omega_m}=0.315\pm0.017$ and $H_0=67.3\pm1.2 \; \rm{kms}^{-1} \;
\rm{Mpc}^{-1}$ for the flat $\Lambda$CDM model, which is in perfect
agreement with our standard ruler result. In contrast, compared with
our quasar sample, recent combined SNLS SNe Ia data favors a lower
value of $\Omega_m$ and thus smaller matter density in the
$\Lambda$CDM model than our quasar data \citep{Conley11}. Let us
note that the cosmological probe inferred from CMB anisotropy
measured by \textit{Planck} is also a standard ruler --- the
comoving size of the acoustic horizon. Therefore appreciable
consistency between the same type of probes (standard rulers) could
be expected and indeed is revealed here.

\begin{figure*}
\begin{center}
  \includegraphics[scale=0.4]{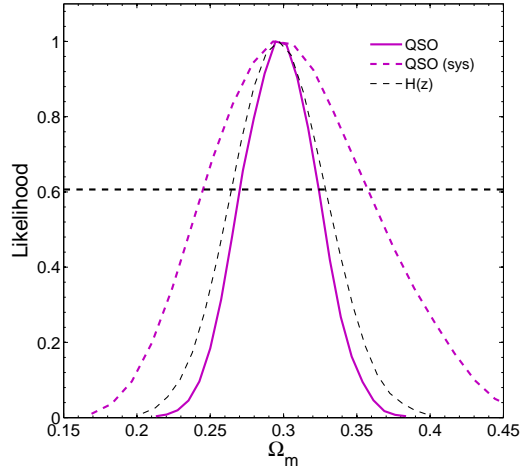}
  \end{center}
  \caption{ The probability distribution of the matter density parameter in the flat $\Lambda$CDM
  model, which is derived from the quasar sample without (red solid line) and with systematical uncertainties (red dashed line) with the
prior on the Hubble constant $H_{0}=70$ km $\rm s^{-1}$ $\rm
Mpc^{-1}$. Fitting result from recent $H(z)$ measurements (black
dashed lines) is also added for comparison. }\label{LCDM2}
\end{figure*}

We emphasize that the value of the Hubble constant obtained in our
analysis, is in excellent agreement with the findings based on
\textit{Planck} CMB data. Many previous have determined its present
value with other probes. For example, the final results of the
Hubble Space Telescope (\textit{HST}) key project suggested the
Hubble constant as $H_0=72\pm8 \; \rm{kms}^{-1}\rm{Mpc}^{-1}$
\citep{Freedman01}. Then, the observations of 240 \textit{HST}
Galactic Cepheid variables gave $H_0=74.2\pm3.6 \;
\rm{kms}^{-1}\rm{Mpc}^{-1}$ \citep{Riess09}. Much lower value has
been suggested by \citet{Tammann08} who independently calibrated
Cepheids and SN Ia and obtained $H_0=62.3\pm1.3 \;
\rm{kms}^{-1}\rm{Mpc}^{-1}$. Two most recent measurements of the
Hubble constant are $H_{0}=69.6\pm0.7$ km $\rm s^{-1}$ $\rm
Mpc^{-1}$ \citep{Bennett14} and that obtained from local Cepheids
distance ladder, $H_0=73.24\pm1.74 \; \rm{kms}^{-1}\rm{Mpc}^{-1}$
\citep{Riess16}. It is also worth noticing that, according to the
meta-analysis of existing literature based on median statistics
\citep{Gott01,Chen03} the value of $H_0=68 \; \rm{kms}^{-1} \;
\rm{Mpc}^{-1}$ can be considered as the most likely value for the
Hubble constant. In order to check the cosmological constraint power
of the quasar sample derived in this analysis, we set the $H_{0}=70$
km $\rm s^{-1}$ $\rm Mpc^{-1}$ prior and obtain a very stringent fit
on the the present-day matter density $\Omega_m=0.297\pm0.027$
(without systematics) and $\Omega_m=0.300\pm0.055$ (with
systematics). This is shown in Fig.~\ref{LCDM2}. For comparison,
fitting result from the $H(z)$ data is also plotted with black
dashed line. It is obvious that the current quasar observations
could provide consistent and comparable cosmological constraints
with respect to the cosmic chronometers.

\subsection{XCDM model}

\begin{figure*}
\begin{center}
  \includegraphics[width=7cm,angle=0]{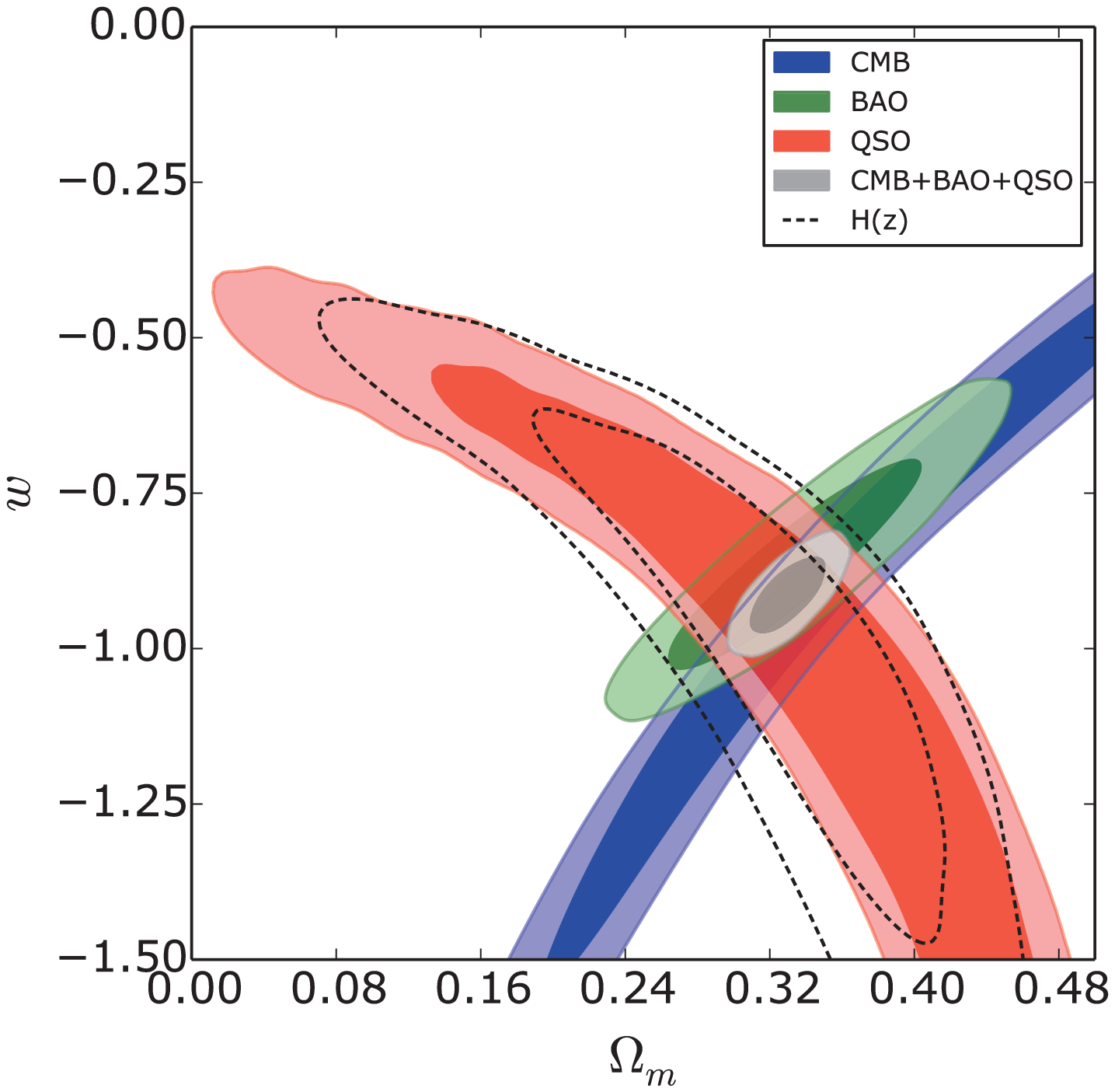} \includegraphics[width=7cm,angle=0]{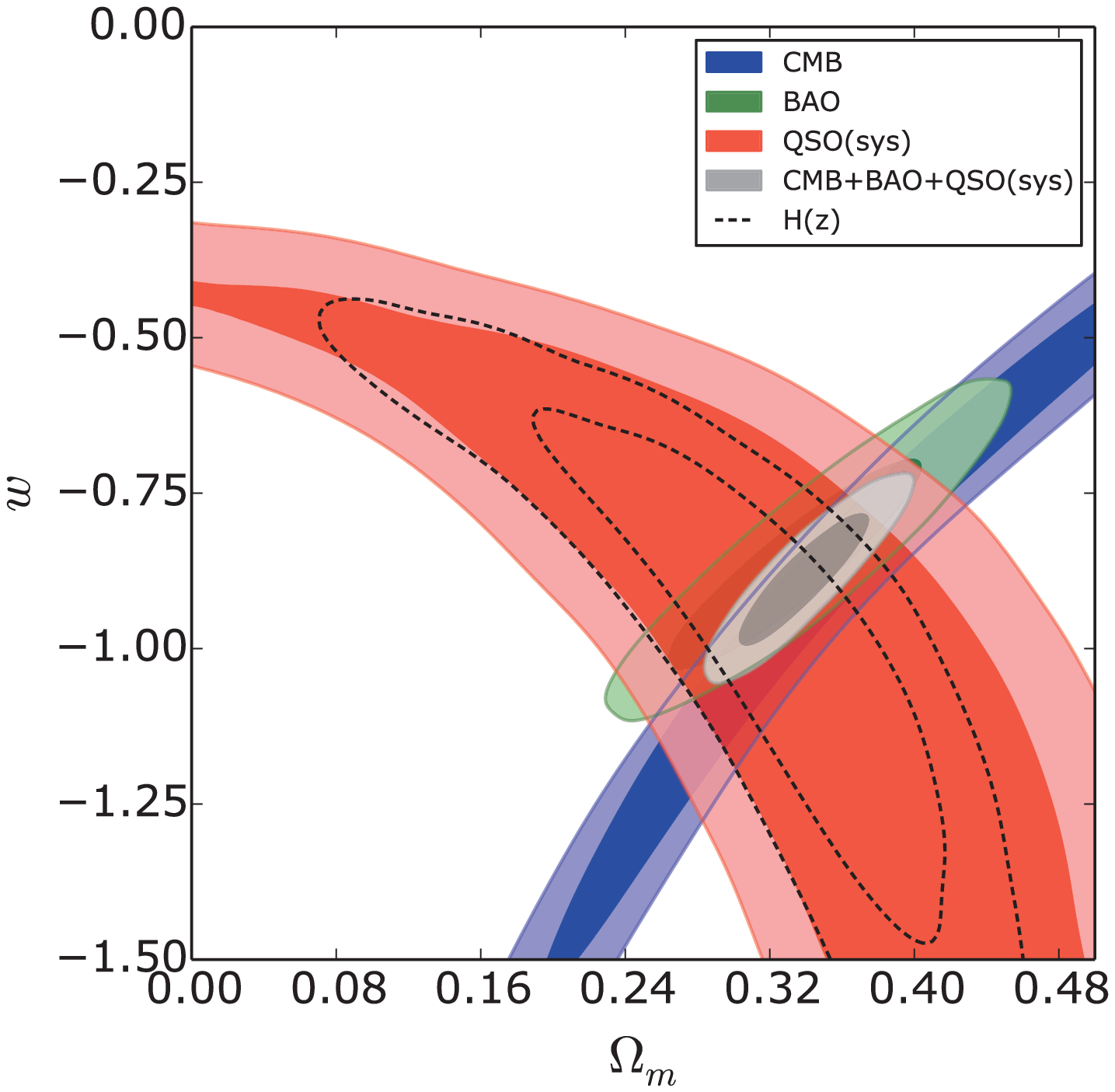}
  \end{center}
\caption{ XCDM model: 68.3\% and 95.4\% confidence regions in the
($\Omega_m$, $w$) plane from QSO, BAO, CMB and $H(z)$. The left
panel shows the QSO confidence region with the best-fit value for
$l$ only, while the right panel shows the confidence region
including systematical uncertainties of $l$, $\beta$ and $n$. We
note that CMB and QSO constraints are orthogonal, making this
combination of cosmological probes very powerful for investigating
the nature of dark energy. }\label{XCDM}
\end{figure*}

Allowing for a deviation from the simple $w=-1$ case, an alternative
is dynamical energy based exclusively on a scalar field
\citep{Ratra98}. In this case, accelerated expansion is obtained
when $w<-1/3$, while scalar field models typically have time varying
$w$ with $w\geq-1$. When flatness is assumed, it is a two-parameter
model with the parameter set: $\textbf{p}=\{\Omega_m, w\}$. From the
fitting results shown in Table.~2 one can see that
${\Omega_m}=0.309^{+0.215}_{-0.151}$, $w=-0.970^{+0.500}_{-1.730}$
(without systematics) and ${\Omega_m}=0.295^{+0.213}_{-0.157}$,
$w=-1.130^{+0.630}_{-2.120}$ (with systematics). In order to
illustrate the performance of QSO data, we also present the
constraints resulting from $H(z)$ data, which clearly indicates that
the current quasar observations confronted with the cosmic
chronometers could provide consistent and comparable cosmological
constraints. In particular, the $w$ coefficient obtained from our
quasar sample agrees very well with the respective value derived
from the \textit{Planck} results. Fig.~\ref{XCDM} shows the contours
for $\Omega_m$ and $w$, with and without systematical uncertainties.
It can be seen that the concordance $\Lambda$CDM model ($w=-1$), is
consistent with quasar method applied here. Our results demonstrate
that the method extensively investigated in our work on
observational radio quasar data can be used in practice to
effectively derive cosmological information.

Angular diameter distances for intermediate-luminosity radio quasars
obtained using the method described in this paper may also
contribute to testing the consistency between luminosity and angular
diameter distances known as distance duality relation. Recent
discussions of the Etherington reciprocity relation, can be found in
\citet{Cao11a,Cao11b,Cao14}. Concerning the latest Union2.1
compilation comprising 580 SN Ia data points \citep{Suzuki12}, the
results obtained from our quasar sample are fully consistent with
the SNIa fits: $\Omega_m=0.296^{+0.102}_{-0.180}$,
$w=-1.001^{+0.348}_{-0.398}$. More importantly, from the comparison
between Fig.~6 in \citet{Suzuki12} and Fig.~\ref{XCDM} in this
paper, one can clearly see that principal axes of confidence regions
obtained with SNe and quasars are inclined at very high angles. This
creates hopes for more stringent constraints in combined analysis of
these two data sets. Considering the big difference between the
sample sizes of the two data sets, we hope intermediate-luminosity
quasars would eventually serve as a complementary probe breaking the
degeneracy in the ($\Omega_m, w$) plane at much higher redshifts.

It is now crucial to pin down the uncertainties of each approach and
employ multiple independent probes to account for unknown
systematics. For comparison, we also plot the likelihood contours
with the latest measurements of BAO and CMB. For the BAO data, we
use the latest measurements of acoustic-scale distance ratio
$D_V(z)/r_s(z_{d})$ from the 6dFGS
($r_s(z_d)/D_V(z=0.106)=0.336\pm0.015$), SDSS-MGS
($D_V(z=0.15)/r_s(z_d)=(664\pm25)/152.66$), BOSS-LOWZ
($D_V(z=0.32)/r_s(z_d)=(1264\pm25)/153.19$) and BOSS-CMASS
($D_V(z=0.57)/r_s(z_d)=(2056\pm20)/153.19$)
\citep{Beutler11,Ross15,Anderson14} \footnote{As discussed in
\citet{Ade16a}, because the WiggleZ volume partially overlaps that
of the BOSS-CMASS sample and the correlations have not been
quantified, we choose not to use the recent WiggleZ results in our
analysis.}. $D_V(z)$ in the distance ratio is the volume-averaged
effective distance defined as
\begin{equation}
D_V(z)=\left[(1+z)^2D_A^2(z)\frac{cz}{H(z)}\right]^{1/3}.
\end{equation} and $r_s(z_{d})$ is the comoving sound horizon
\begin{equation}
r_s(z_{d}) ={H_0}^{-1}\int_{z_{d}}^{\infty}c_s(z)/E(z')dz'.
\end{equation}
at the baryon-drag epoch, $z_{d}$, which can be calculated as
\citep{Eisenstein98}
\begin{equation}
z_d=\frac{1291(\Omega_{m}h^2)^{0.251}}{1+0.659(\Omega_{m}h^2)^{0.828}}[1+b_1(\Omega_{b}h^2)^{b_2}]
\end{equation}
where
$b_1=0.313(\Omega_{m}h^2)^{-0.419}[1+0.607(\Omega_{m}h^2)^{0.674})$
and $b_2=0.238(\Omega_{m}h^2)^{0.223}$. For the CMB data, we use the
distance priors derived from the recent $Planck$ data \citep{Ade16},
which include the measurements of of the derived quantities, such as
the acoustic scale ($l_A$), the shift parameter ($\mathcal{R}$), and
the baryonic fraction parameter ($\Omega_{b}h^2$). The acoustic
scale at recombination can be parametrized as
\begin{equation}
l_A\equiv(1+z_*)\frac{\pi D_A(z_*)}{r_s(z_*)}
\end{equation}
where the comoving sound horizon expresses as
\begin{equation}
r_s(z)=\int^a_0\frac{da}{a^2E(a)\sqrt{3(1+Ra)}}
\end{equation}
with $R=31500(T_{CMB}/2.7K)^{-4}\Omega_{b}h^2, T_{CMB}=2.7255K$. The
redshift of photo-decoupling period, $z_*$, can be calculated as
\citep{Hu96}
\begin{equation}
z_*=1048[1+0.00124(\Omega_{b}h^2)^{-0.738}][1+g_1(\Omega_{m}h^2)^{g_2}]
\end{equation}
where
$g_1=\frac{0.0783(\Omega_{b}h^2)^{-0.238}}{1+39.5(\Omega_{b}h^2)^{0.763}},g_2=\frac{0.560}{1+21.1(\Omega_{b}h^2)^{1.81}}$.
The $\mathcal{R}$ quantity is the least cosmological model-dependent
parameter that can be extracted from the analysis of the CMB and
takes the form
\begin{equation}
\mathcal{R}(z_*)\equiv\frac{(1+z_*)D_A(z_*)\sqrt{\Omega_{m}H_0^2}}{c}
\end{equation}
Combined with the inverse covariance matrix $C^{-1}_{CMB}$ from
\citet{Ade16}, the contribution of CMB to the $\chi^2$ value can be
written as
\begin{equation}
\chi^2_{CMB}=\Delta{P}^T_{CMB}C^{-1}_{CMB}\Delta{P}_{CMB}
\end{equation}
where $\Delta{P}_{CMB}$ is the difference between the theoretical
distance prior and the observational one. In Fig.~\ref{XCDM}, we
show the confidence contours of $\Omega_m$ and $w$ from QSO, BAO and
CMB. Both the individual constraints and the combined constraint are
shown. The QSO constraint is almost orthogonal to that of the CMB
and BAO. Adding the constraints from BAO and CMB reduces the
uncertainty. Under the assumption of a flat Universe, the three
probes together yield $\Omega_m=0.331^{+0.022}_{-0.022}$,
$w=-0.939^{+0.075}_{-0.075}$ and $\Omega_m=0.331^{+0.042}_{-0.035}$,
$w=-0.937^{+0.139}_{-0.134}$ without and with systematical
uncertainties, respectively. From Fig.~\ref{XCDM} and Table 2, it is
easy to see that the combined angular diameter distance data favors
a slightly larger values of both $\Omega_m$ and $w$, while QSOs data
favors a relatively larger $\Omega_m$ and a smaller $w$.

\subsection{Time Dependent Equation of State}

Next, we examined models with the DE equation of state allowed to
vary with time. Considering that the quasar data alone do not
tightly constrain $w$, even for spatially flat models, more data are
added to break the strong geometrical degeneracy.

\begin{figure*}
\begin{center}
  \includegraphics[width=8cm,angle=0]{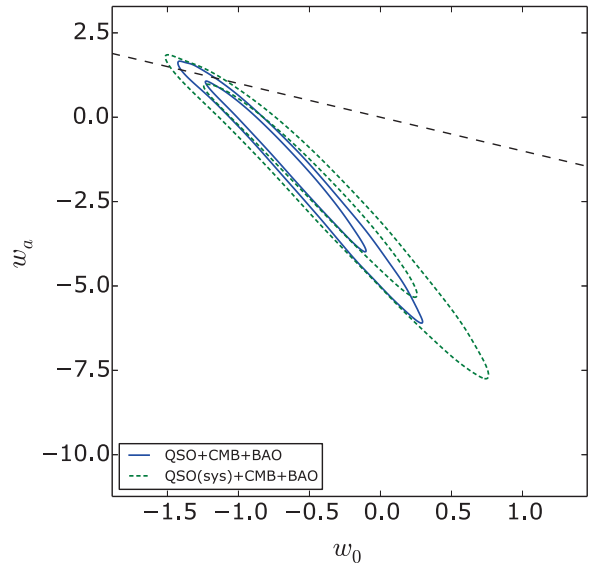}
  \end{center}
\caption{ 68.3\% and 95.4\% confidence regions of the $(w_0,w_a)$
plane from QSO, BAO, CMB, and their combination. Zero curvature has
been assumed. Points above the dotted line ($w_0+w_a>0$) violate
early matter domination and are disfavored by the data. }\label{CPL}
\end{figure*}

Among a wide range of dark energy models, we consider the commonly
used Chevalier-Polarski-Linder (CPL) model involving certain
dynamical scalar field models \citep{Chevalier01,Linder03}, in
which, to good approximation, the the equation of state of dark
energy is parameterized as
\begin{equation}
w(z)=w_{0}+w_{a}z/(1+z)
\end{equation}
where $w_0$ and $w_a$ are constants and the $\Lambda$CDM model is
recovered when $w_{0}=-1$ and $w_a=0$. Adding \textit{Planck} CMB
and BAO and to the quasar data gives the 68.3\% constraints:
$\Omega_m=0.320^{+0.029}_{-0.022}$, $w_0=-0.606^{+0.578}_{-0.544}$,
$w_a=-1.207^{+2.335}_{-2.684}$ (without systematics) and
$\Omega_m=0.330^{+0.032}_{-0.035}$, $w_0=-0.439^{+0.830}_{-0.681}$,
$w_a=-1.816^{+2.878}_{-3.493}$ (with systematics). The constraints
on $w_{0}$ and $w_{a}$ are shown in Fig.~\ref{CPL} and Table 2. Note
that the combined angular diameter distance data favors a $w_0>-1$
and a negative $w_a$, which means that dark energy was phantom-like
($w<-1$) in the past, then its EoS crossed the phantom divide, and
became quintessence-like ($w>-1$) recently; finally its EoS will
become positive in the future. On the contrary, QSO data with $l$
prior favors a $w_0<-1$ and a positive $w_a$, which means that dark
energy was quintessence-like ($w>-1$) in the past, then its EoS
crossed the phantom divide, and became phantom-like ($w<-1$)
recently; finally the universe will end in a big rip.

\begin{figure*}
\begin{center}
  \includegraphics[width=8cm,angle=0]{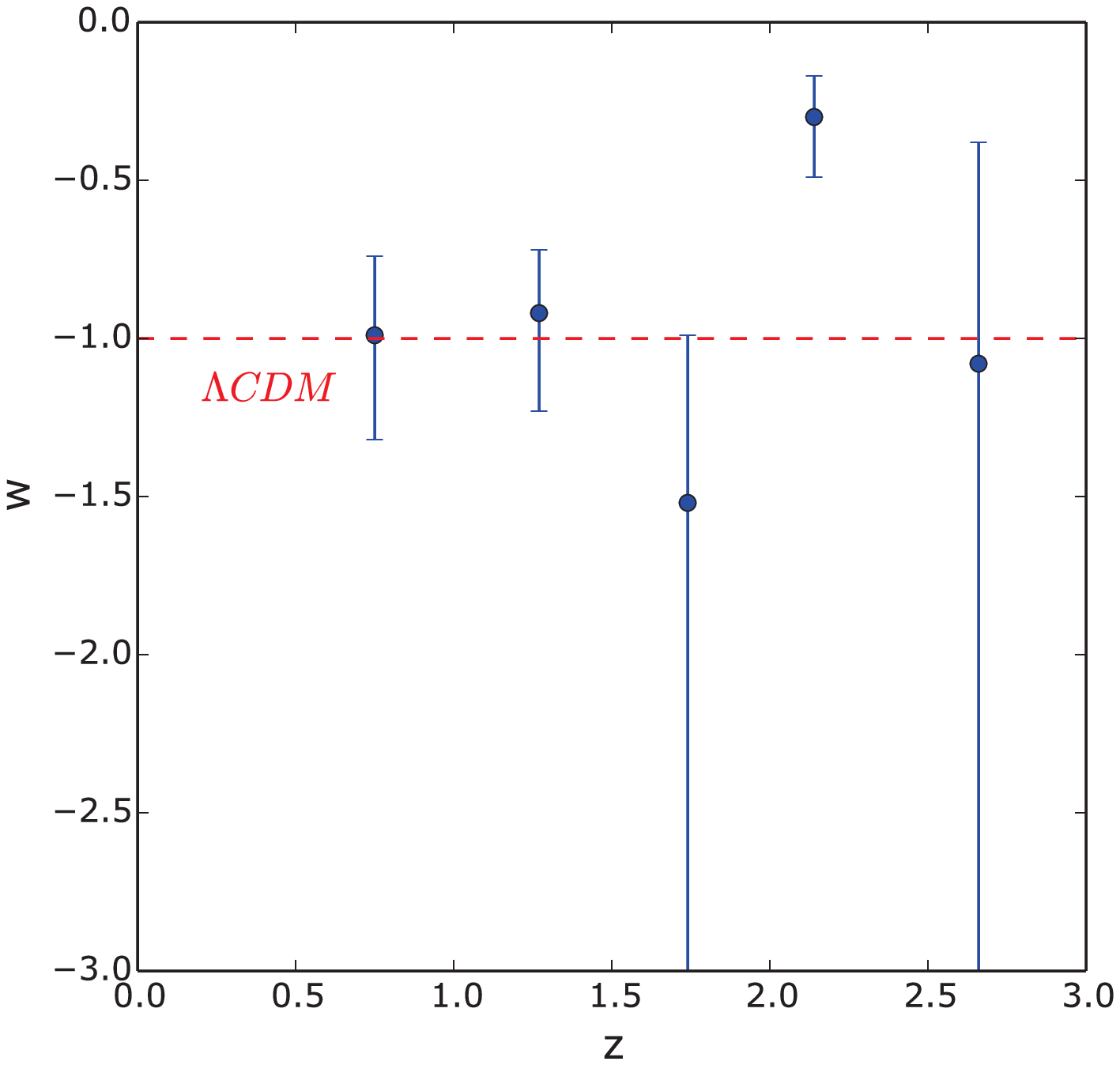} \includegraphics[width=8cm,angle=0]{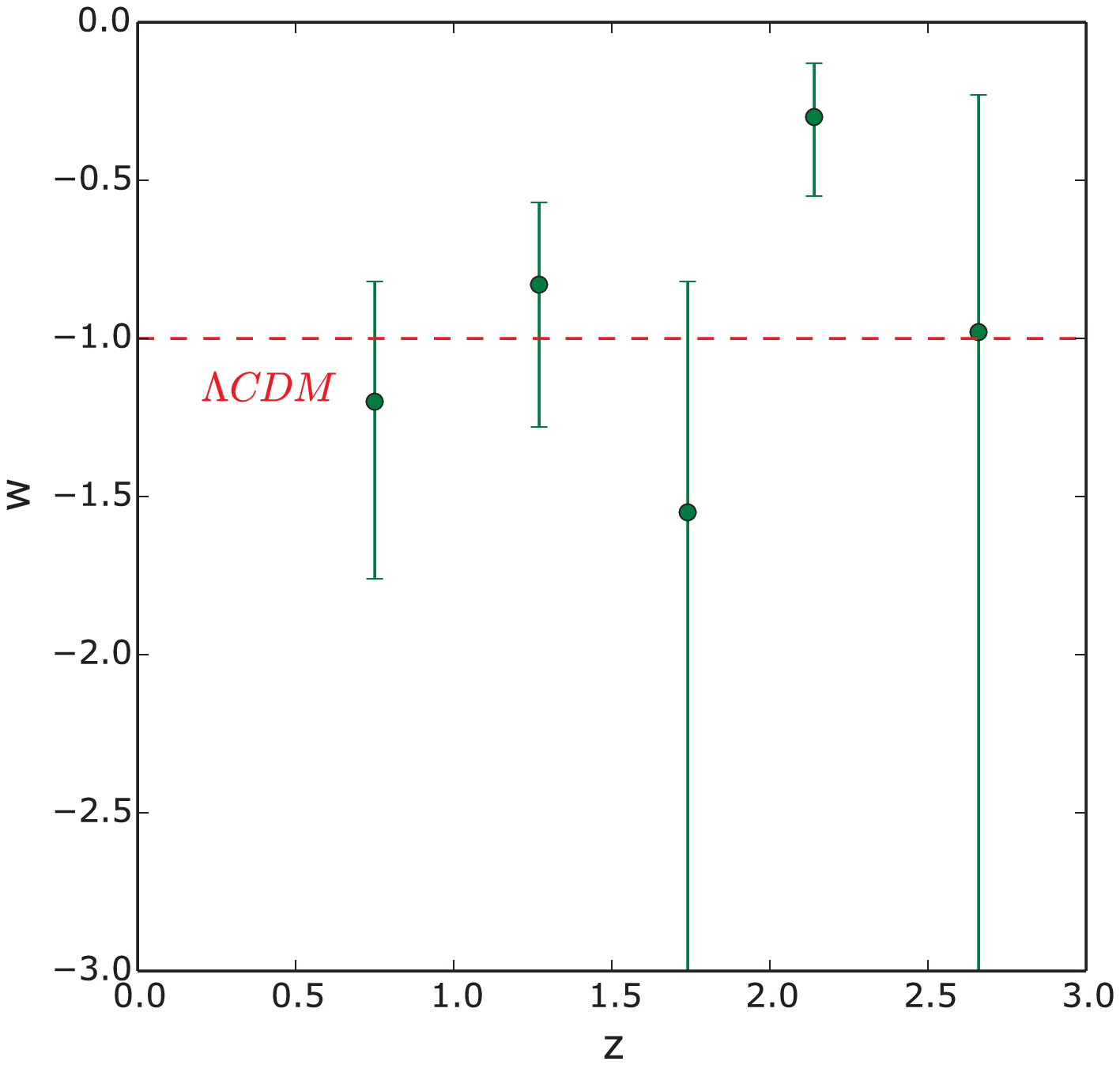}
  \end{center}
\caption{ The best fitted $w$ with marginalized $68.3\%$ CL error.
The results were obtained assuming a flat universe for the joint
data set of QSO, BAO, CMB, with (left panel) and without (right
panel) QSO systematics. The QSOs are divided into five groups with
$z<1.0$, $1.0<z<1.5$, $1.5<z<2.0$, $2.0<z<2.5$ and $z>2.5$, while
BAO and CMB are added to constrain the value of $\Omega_m$,
considering the well known ``geometrical degeneracy'' problem. We
emphasize that the results are still consistent with the
cosmological constant (red-dashed line) at the 68.3\% confidence
level. }\label{subsample}
\end{figure*}

An accurate reconstruction of $w(z)$ can considerably improve our
understanding the nature of both dark energy and gravity. In order
to reconstruct the evolution of $w(z)$ without assuming a specific
form, such model will inevitably include more parameters than
$w_0-w_a$, the number of dark-energy equation-of-state parameters
depending on the number of redshift bins \citep{Kowalski08}.
Confined to the sample size of our available quasar data, we carry
out the analysis by dividing the full sample into different
sub-samples given their redshifts and fitting a constant $w$ in each
sub-sample. The redshifts of the QSOs span from $z=0.462$ to
$z=2.73$, so we divide the QSOs into five groups with $z<1.0$,
$1.0<z<1.5$, $1.5<z<2.0$, $2.0<z<2.5$ and $z>2.5$, respectively. The
first group has 30 QSOs with redshifts $z<1.0$, the second group has
51 QSOs with redshifts $1.0<z<1.5$, the third group has 25 QSOs with
redshifts $1.5<z<2.0$, the fourth group has 11 QSOs with redshifts
$2.0<z<2.5$ and the fifth group contains 3 QSOs. We then fit the
cosmic equation of state to each group of QSOs, while the remaining
cosmological parameters are fixed at the best-fit values determined
by Planck results. The constraints are shown in Fig.~\ref{subsample}
and Table 4.

The first group shows a well-constrained equation-of-state parameter
from redshift 0.5 to 1.0. Therefore, no evidence of deviation from
$w=-1$ is detected from low-redshift quasars, which is in good
agreement with the previous findings from Union2.1 SN Ia constraints
\citep{Amanullah10}. The deviation from $\Lambda$CDM is also not
obvious in the second and fifth redshift groups. Interestingly, our
quasar data favors a transition from $w<-1$ at low redshift to
$w>-1$ at higher redshift, a behavior that is consistent with the
quintom model allowing $w$ to cross -1. A redshift bin shifts the
confidence interval for $w(2.0<z<2.5)$ towards higher $w$, which is
typically favored by many scalar field models. As shown in
Fig.~\ref{subsample}, the transition redshift at which $w$ departing
from -1 is apparently located at $z\sim 2.0$, which might be
overlooked by the recent analysis with a joint data set including
Union2.1 SN, CMB, $H(z)$, RSD (redshift space distortion) and BAO
while fixing $w=-1$ at $z>1.5$ \citep{Zhao12}. More data extending
above redshift $z=3$ will be necessary to investigate the dark
energy equation-of state parameter in this high redshift region
where the uncertainty is still very large.

\section{Beyond dark energy} \label{Alternative}

As is well known, a physically profound question to be addressed in
the empirical cosmological studies is: does the cosmic acceleration
arise from a new energy component with repulsive gravity or a
breakdown of General Relativity (GR) on cosmological scales? In this
section, we will investigate the constraining power of our quasar
sample, concerning different approaches to explain the accelerated
expansion of the Universe based on the departure from classical GR.
Here we lay the framework for such options and give some examples.
In particular we will consider:
\begin{itemize}
\item[] RDE: Ricci dark energy model in a flat universe.
\item[] DGP: Dvali-Gabadadze-Porrati brane world model in a flat universe.
\end{itemize}

\subsection{Ricci dark energy}

\begin{figure*}
\begin{center}
  \includegraphics[width=8cm,angle=0]{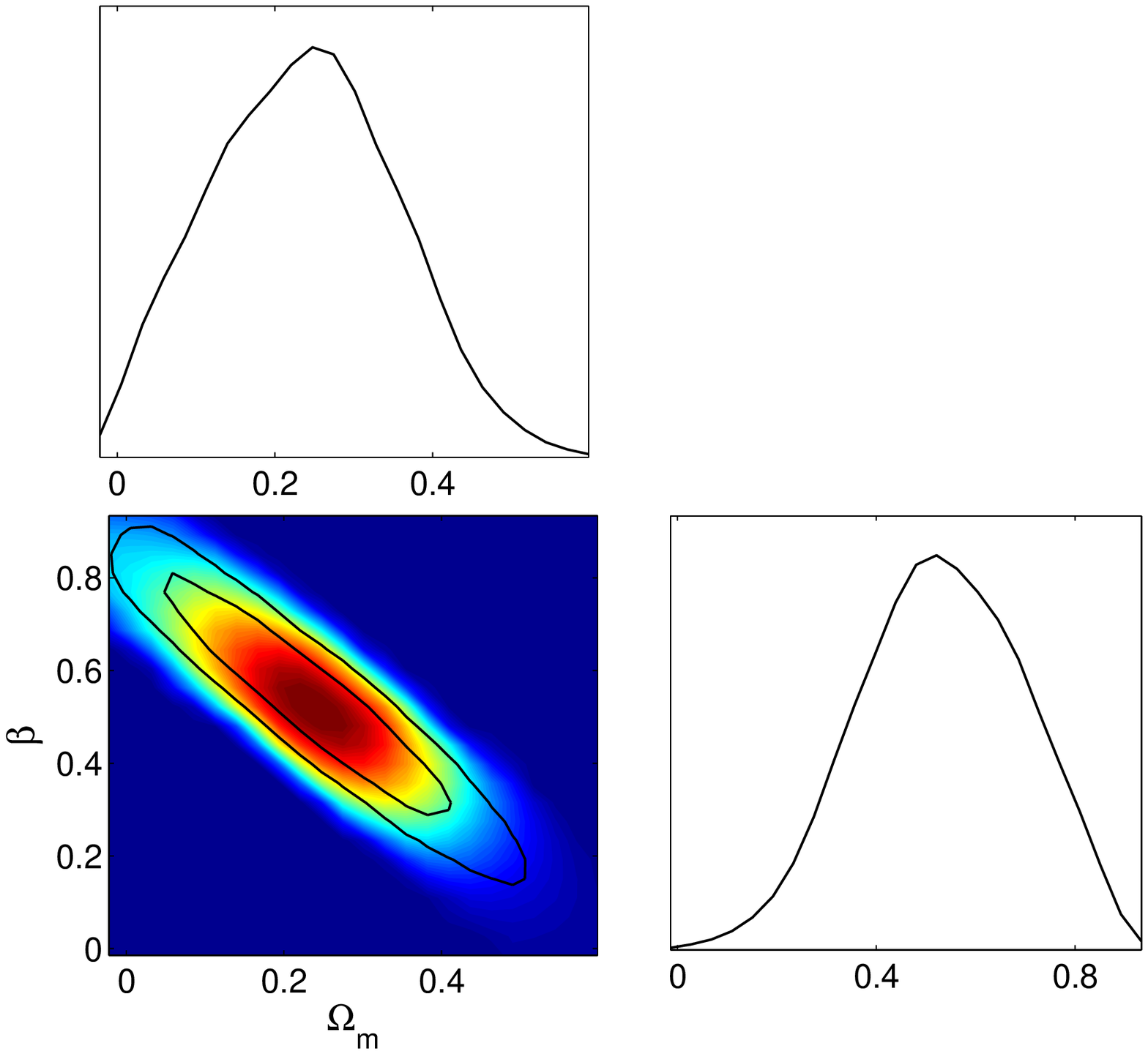} \includegraphics[width=8cm,angle=0]{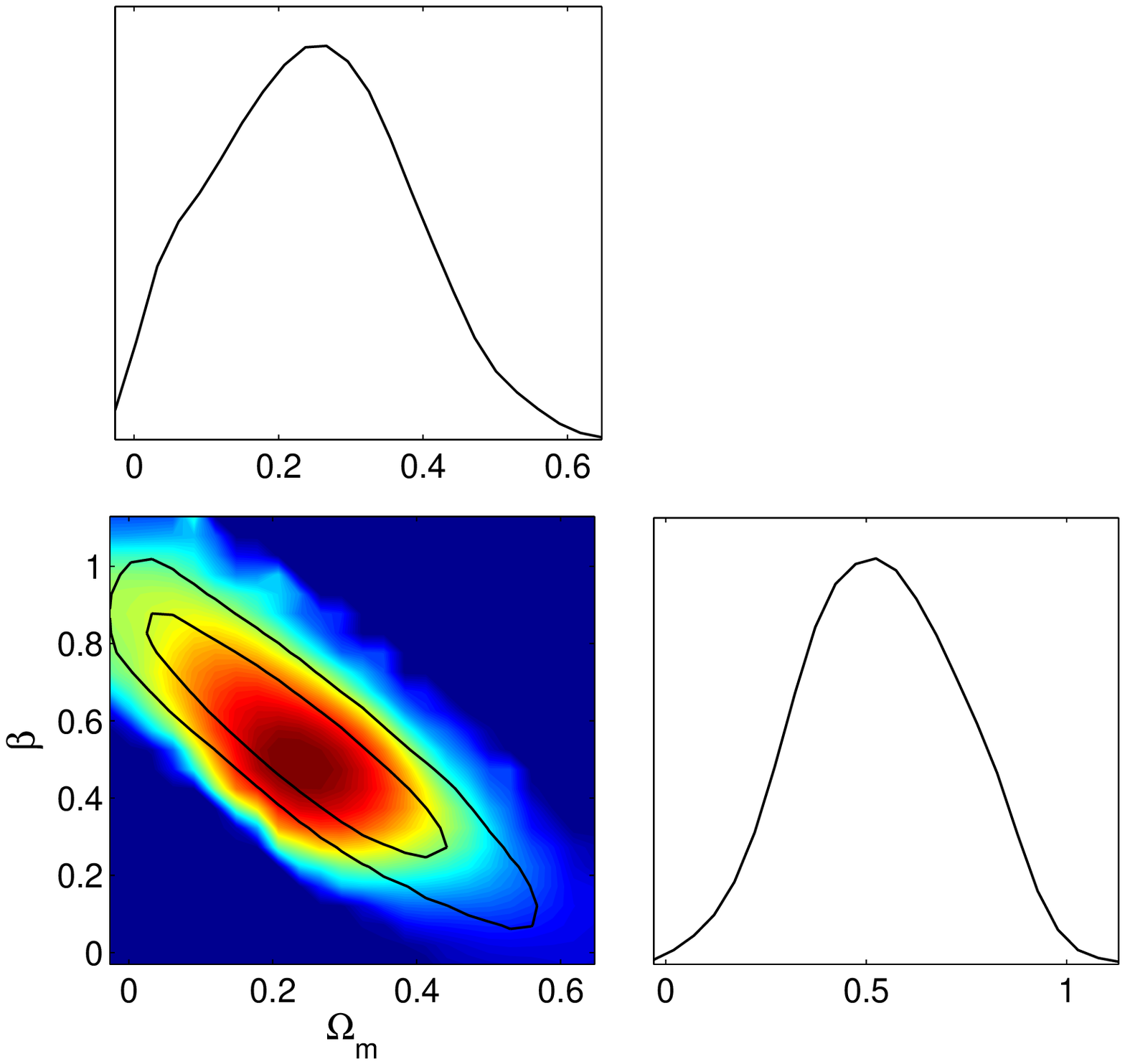}
  \end{center}
\caption{ The 2-D regions and 1-D marginalized distribution of RDE
model parameters from the quasar sample, without (left panel) and
with systematical uncertainties (right panel). }\label{Racci}
\end{figure*}

Other cosmological approaches to describe the dark component have
received considerable attention in the past, one of which is
holographic dark energy, proposed in the context of the fundamental
principle of quantum gravity \citep{Bekenstein81,Gao09}. Compared
with the cosmological constant model, this mechanism may alleviate
the well-known coincidence problem and fine tuning problem. The
idea, here is that the scale of dark energy is set by a cosmological
Hubble horizon scale instead of the Planck length. Choosing an
Infrared (IR) cutoff of the quantum field theory as $|R|^{-1/2}$,
where $R = 6(\dot{H}+2H^2)$ is the Ricci scalar of the flat
Friedman-Robertson-Walker metric, one can derive an effective
equivalent of the dark energy density \citep{Gao09}
 \begin{equation}
 {\rho_{de}}={3{\beta}M^2_{Pl}(\dot{H}+2H^2)}
 \end{equation}
where $\beta$ is a constant parameter larger than zero. The Hubble
parameter can be derived from the Friedman equation:
\begin{equation}
H^2 = H_0^2 \left[\frac{
2\Omega_{m}}{2-\beta}(1+z)^{3}+(1-{2\Omega_{m}\over 2
-\beta})(1+z)^{(4-{2\over\beta})}\right] \, .
\end{equation}
This is a two-parameter model with $\textbf{p}=\{\Omega_m,~\beta\}$.
Testing the RDE model with the quasar data, we obtain the following
best fits: ${\Omega_m}=0.229^{+0.184}_{-0.184}$,
$\beta=0.550^{+0.265}_{-0.265}$ (without systematics) and
${\Omega_m}=0.240^{+0.210}_{-0.210}$,
$\beta=0.520^{+0.365}_{-0.275}$ (with systematics). These results
shown presented in Fig.~\ref{Racci} and Table~4 are in agreement
with the previous analysis using galactic-scale strong gravitational
lensing systems \citep{Biesiada11,Cao12c}, as well as the previous
work based on the SNe Ia Constitution compilation, the BAO
measurement from the SDSS and the Two Degree Field Galaxy Redshift
Survey, and the CMB measurement given by the five-year WMAP
observations \citep{Li09}.

\subsection{Higher dimension theories}

\begin{figure*}
\begin{center}
  \includegraphics[width=8cm,angle=0]{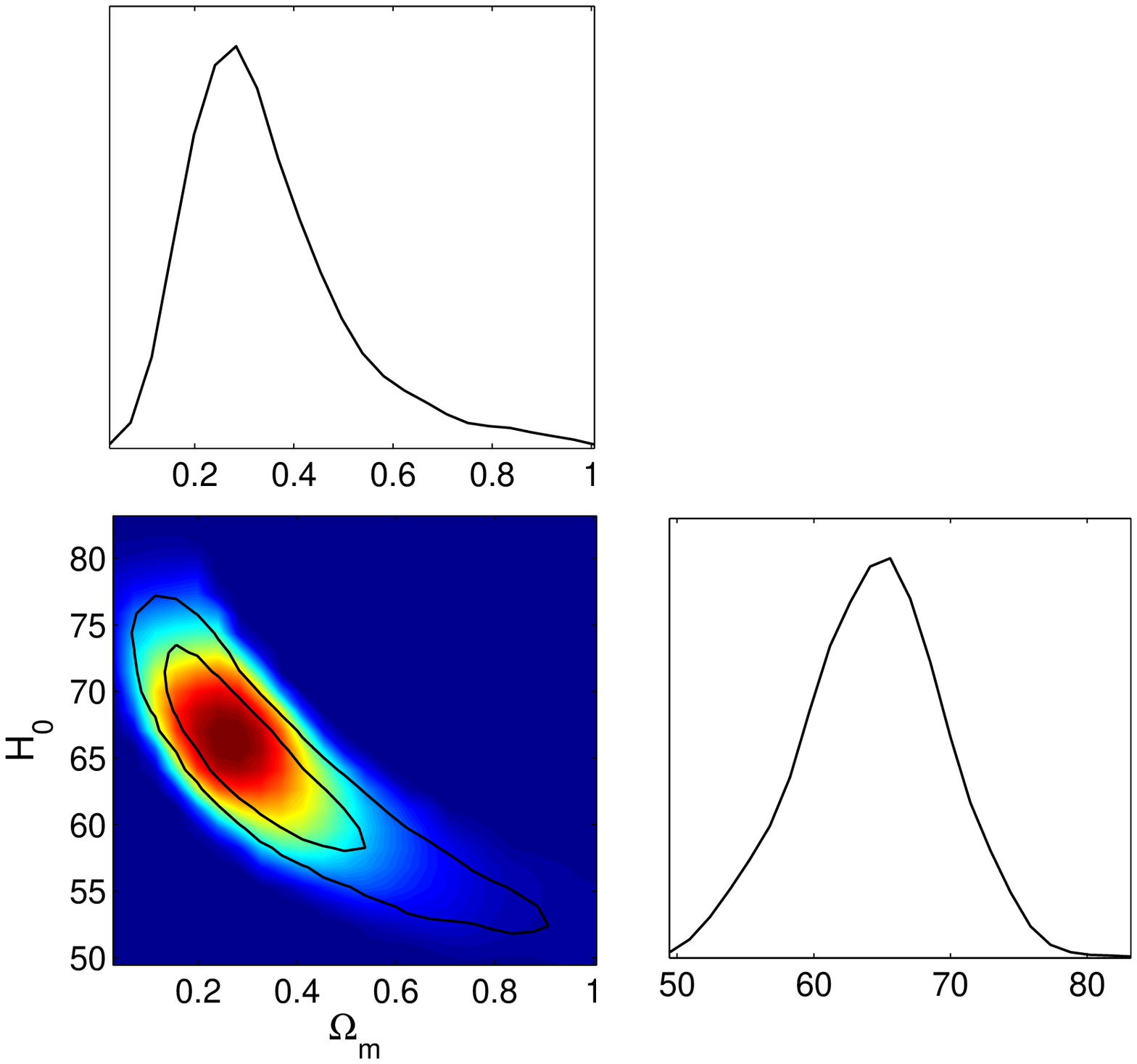} \includegraphics[width=8cm,angle=0]{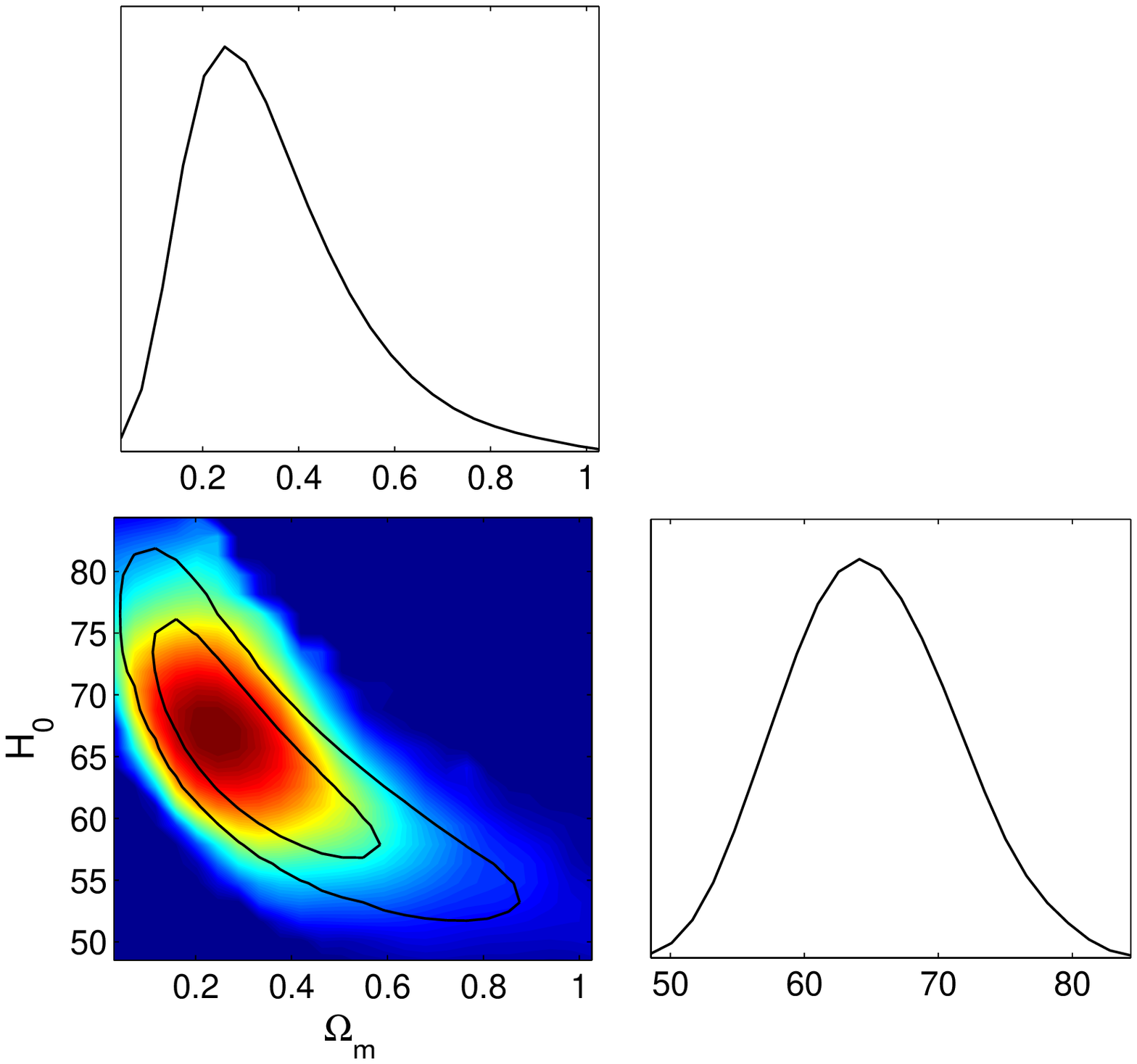}
  \end{center}
\caption{ The 2-D regions and 1-D marginalized distribution of DGP
model parameters from the quasar sample, without (left panel) and
with systematical uncertainties (right panel). }\label{DGP}
\end{figure*}

Past decades have witnessed considerable advances in the
modification of General Relativity, as a possible explanation of
accelerated expansion of the Universe. One radical proposal is to
introduce extra dimensions and allow gravitons to leak off the brane
representing the observable universe. Embedding our 4-dimensional
spacetime into a higher dimensional bulk spacetime, \citet{Dvali00}
proposed the well-known Dvali-Gabadadze-Porrati (DGP) brane world
model, in which the leaking of gravity above a certain cosmological
scale $r_c$ might be responsible for the increasing cosmic expansion
rate. Inspired by the DGP example, a general class of ``galileon''
and massive gravity models has been proposed in the literature
\citep{Mortonson14}. The length at which gravity leaking occurs
defines an omega parameter: $\Omega_{r_c}=1/(4r^2_cH^2_0)$, with
which the Friedman equation modified as
\begin{equation}
H^2 = H_0^2
(\sqrt{\Omega_{m}(1+z)^3+\Omega_{r_c}}+\sqrt{\Omega_{r_c}})^2,
\end{equation}
The flat DGP model only contains one free model parameter,
${\theta}=\{\Omega_m\}$, which is related to $\Omega_{r_c} =
\frac{1}{4} (1 - \Omega_m)^2$ under assumption of a flat Universe.
In order to make a comparison with the $\Lambda$CDM model, we also
take the Hubble constant as a free parameter. The best fit value for
the mass density parameter in DGP model is
$\Omega_m=0.285^{+0.255}_{-0.155}$ (without systematics) and
$\Omega_m=0.248^{+0.335}_{-0.130}$ (with systematics), while the
best-fit Hubble constant is
$H_0=66.2^{+7.4}_{-8.2}\rm{kms}^{-1}\rm{Mpc}^{-1}$ (without
systematics) and $H_0=64.3^{+11.8}_{-7.6}\rm{kms}^{-1}\rm{Mpc}^{-1}$
(with systematics). These results are presented in Fig.~\ref{DGP}
and Table~4. The DGP model is the one for which we obtained the
tightest constraints in our analysis. This is due to the simplicity
of the model, which depends on only one parameter. In this respect,
this is the simplest model, together with the standard flat
$\Lambda$CDM. In general, as we already discussed in \citep{Cao14},
it is to be expected that models with fewer parameters perform
better.

From the above considerations, two crucial consequences arise:
first, given the current status of cosmological observations
including QSOs, there is no strong reason to go beyond the simple,
standard cosmological model with zero curvature and a cosmological
constant; second, a low value of the Hubble constant is preferred by
both the new \textit{Planck} data and our quasar observations. This
consistency between fundamental cosmological parameters constrained
from the high redshift CMB measurements $z\sim 1000$ and those from
the observations at relatively low redshifts $z\sim 3$ may alleviate
the tension between \textit{Planck} and the SN Ia observations at
$z=0$ \citep{Marra13,Xia13,Li13}. In fact, projected parameters
should presumably be the same from measurements at all $z$ in a
given model.

\section{Discussion: Improving cosmological constraints by efficiently adding low-redshift clusters} \label{Discusssion}

The redshift of intermediate-luminosity quasars ranges between
$z=0.46$ and $z=2.73$. Therefore, we also added to the data a set of
25 well-measured angular diameter distances from the galaxy
clusters. They have been obtained by considering Sunyaev-Zeldovich
effect (SZE) together with X-ray emission of galaxy clusters
\citep{Filippis05}, where an isothermal elliptical $\beta$ model was
used to describe the clusters. The enlargement of $D_A(z)$ sample at
lower redshifts ($0.142<z<0.890$) improves the assessment of the $w$
parameter that describes the properties of dark energy.

As shown in Fig.~\ref{XCDM}, adding the cluster data tightens the
constraints substantially, giving the improved fitting results on
the XCDM model: ${\Omega_m}=0.279^{+0.189}_{-0.189}$,
$w=-1.066^{+0.614}_{-0.614}$. Speaking in terms of the figure of
merit (FoM) --- a measure proposed by the Dark Energy Task Force
\citep{Albrecht06}, which is equal to the inverse of the area of the
95\% confidence contour in the parameter plane, we find that this
combined data set improves the constraint on $w$ by 30\%.
Considering the redshift coverage of these two astrophysical probes,
the combination of high-redshift quasars and low-redshift clusters
may provide an important source of angular diameter distances, in
addition to the previously studied probes including strongly
gravitationally lensed systems
\citep{Biesiada10,Biesiada11,Cao12b,Cao15b} or X-ray gas mass
fraction of galaxy clusters \citep{Allen04,Allen08}.

\begin{figure*}
\begin{center}
  \includegraphics[width=8cm,angle=0]{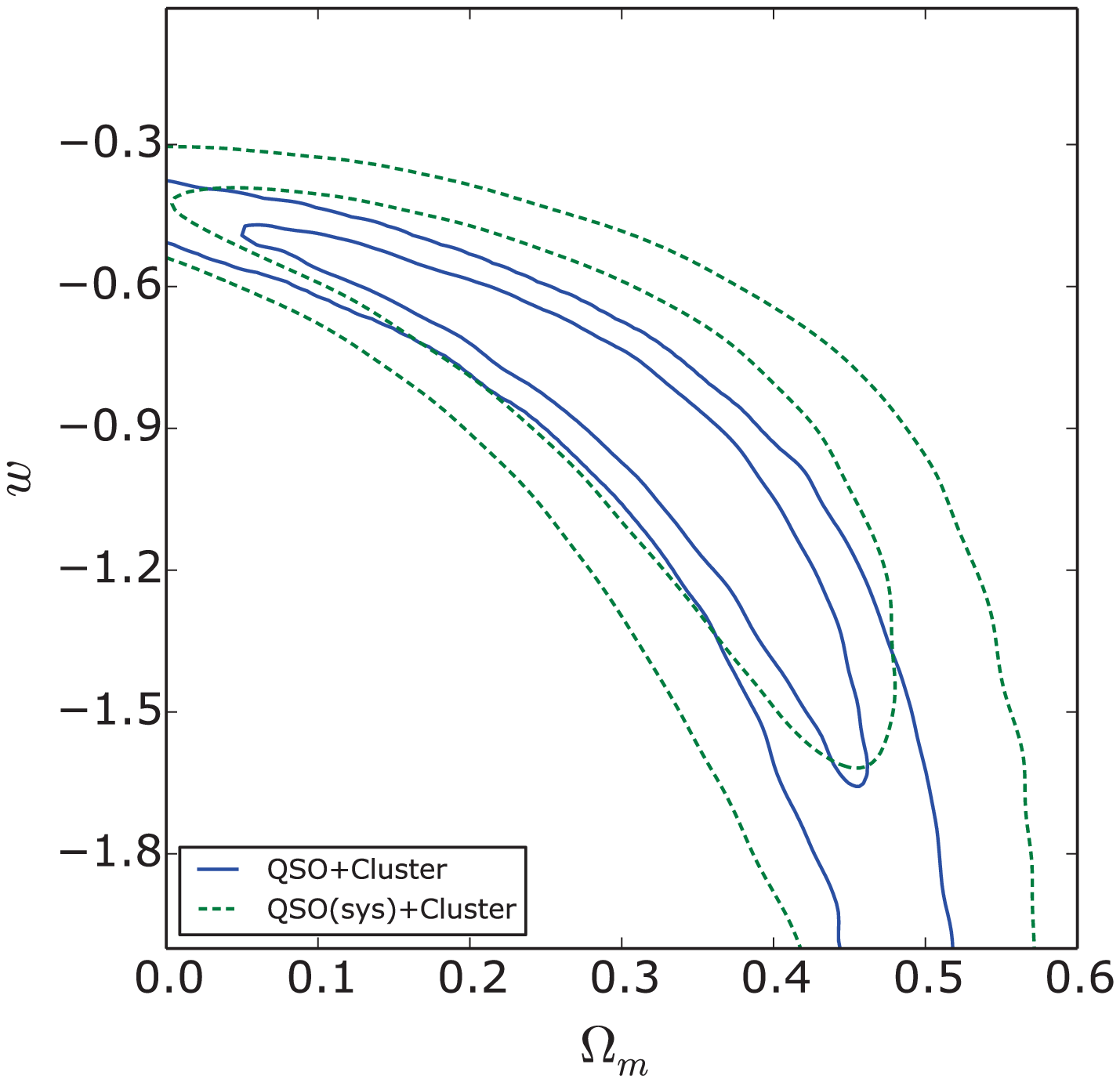} \includegraphics[width=8cm,angle=0]{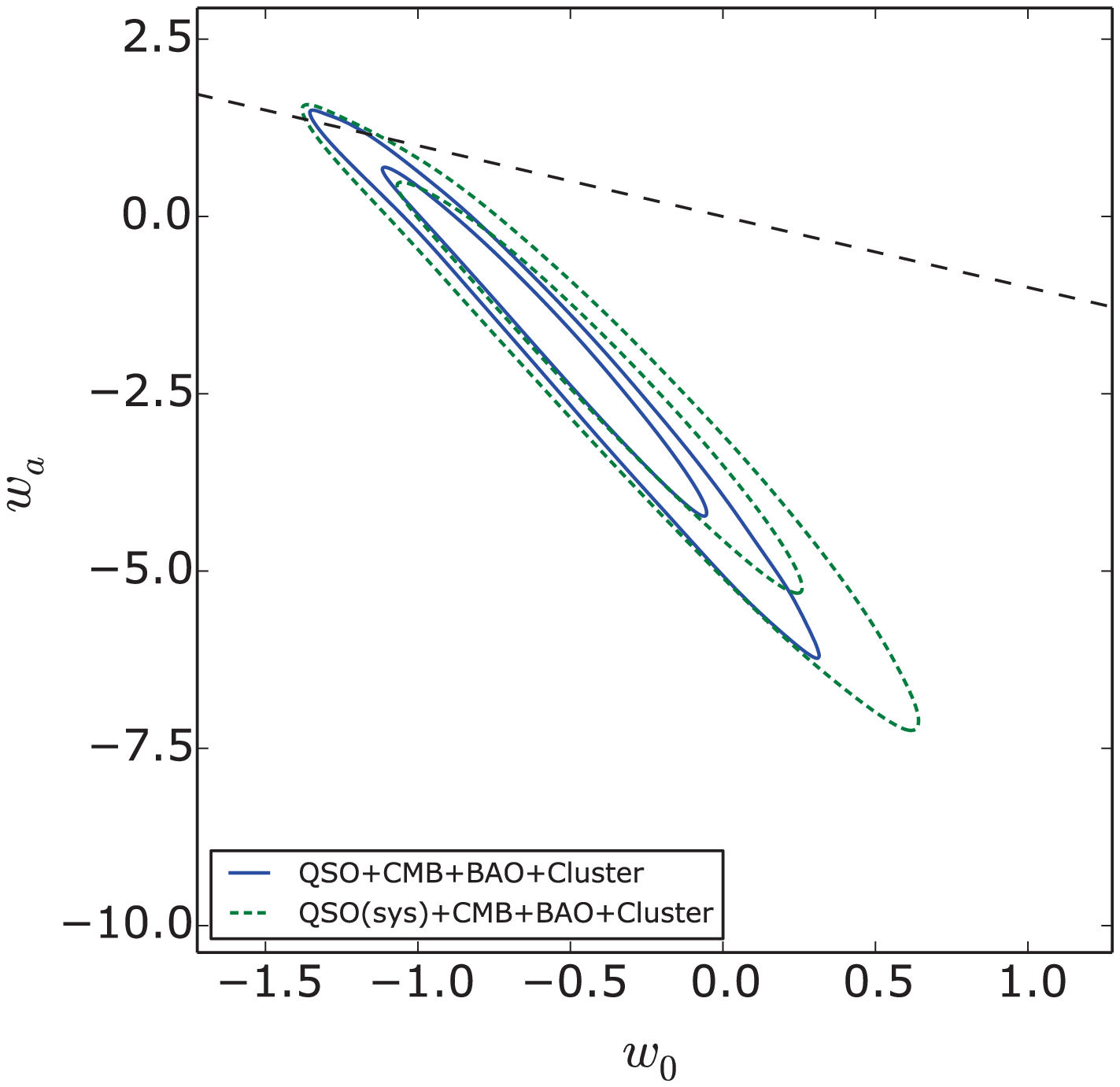}
  \end{center}
\caption{ Best-fit confidence regions in the $\Omega_m$-$w$ and
$w_0-w_a$ plane from quasars combined with angular diameter distance
measurements from cluster observations.
 }\label{cluster}
\end{figure*}

Now we will show how the combination of most recent and
significantly improved cosmological observations can be used to
study the cosmic equation of state. We consider four background
probes which are directly related to angular diameter distances:
intermediate-luminosity quasar data (QSO), Sunyaev-Zeldovich effect
(SZE) together with X-ray emission of galaxy clusters, baryonic
acoustic oscillations (BAO), and CMB observations. The first two
probes are always considered as individual standard rulers while the
other two probes are treated as statistical standard rulers in
cosmology. Results concerning the constraints on the CPL model
parameters are displayed in Fig.~\ref{cluster}, with the best fit
$\Omega_m=0.320^{+0.025}_{-0.022}$, $w_0=-0.532^{+0.488}_{-0.579}$,
$w_a=-1.686^{+2.424}_{-2.506}$ (without systematics) and
$\Omega_m=0.329^{+0.032}_{-0.033}$, $w_0=-0.375^{+0.637}_{-0.690}$,
$w_a=-2.330^{+2.868}_{-2.964}$ (with systematics). At 68.3\% C.L.,
we find that this model is still compatible with $\Lambda$CDM, i.e.
the case ($w_0=-1$; $w_a=0$) typically lies within the 1$\sigma$
boundary. In this context, it is clear that collection of more
complete observational data concerning angular diameter distance
measurements does play a crucial role \citep{Cao14}.

\section{Summary and conclusions} \label{Conclusion}

In this paper, we have presented a newly compiled data set of 120
milliarcsecond compact radio-sources representing
intermediate-luminosity quasars covering the redshift range $0.46< z
<2.76$. These quasars show negligible dependence of their linear
size on the luminosity and redshift ($|n|\simeq 10^{-3}$,
$|\beta|\simeq 10^{-4}$) and thus represents, in the standard model
of cosmology, a fixed comoving-length of standard ruler.
We implemented a new cosmology-independent technique to calibrate
the linear size of this standard ruler. In particular, we used the
technique of Gaussian processes to reconstruct the Hubble function
$H(z)$ as a function of redshift from 15 measurements of the
expansion rate obtained from age estimates of passively evolving
galaxies. This reconstruction enabled us to derive the angular
diameter distance to a certain redshift $z$, and thus calibrate the
liner size of radio quasars. More importantly, we found $l_m=
11.03\pm0.25$ pc is the typical radius at which AGN jets become
opaque at the observed frequency $\nu\sim 2$ GHz. Our measurement of
this linear size is also consistent with both the previous and most
recent radio observations at other different frequencies, in the
framework of the BK79 conical jet model.

Then this new quasar sample was used to investigate the properties
of dark energy.
In the framework of flat $\Lambda$CDM model, a high value of the
matter density parameter, $\Omega_m=0.322^{+0.244}_{-0.141}$, and a
low value of the Hubble constant,
$H_0=67.6^{+7.8}_{-7.4}\;\rm{kms}^{-1}\rm{Mpc}^{-1}$ are obtained,
which is in excellent agreement with the CMB anisotropy measurements
by \textit{Planck}. For the constant $w$ of a dynamical dark-energy
model, we obtained ${\Omega_m}=0.309^{+0.215}_{-0.151}$,
$w=-0.970^{+0.500}_{-1.730}$ at 68.3\% CL, which demonstrates no
significant deviation from the concordance $\Lambda$CDM model.
Consistent fitting results were also derived for other cosmological
mechanisms explaining the cosmic acceleration, including the Ricci
dark energy model and Dvali-Gabadadze-Porrati (DGP) brane world
model. Moreover, we reconstructed the dark-energy equation-of-state
parameter from different quasar sub-sample, and investigated the
evolution of $w$ in the redshift range $0.46< z <2.76$. No evidence
of deviation from $w=-1$ was detected from low-redshift quasars,
which is in good agreement with the previous findings from SN Ia
constraints \citep{Amanullah10,Suzuki12}. Interestingly, the most
likely reconstruction using our quasar data favors the transition
from $w<-1$ at low redshift to $w>-1$ at higher redshift, a behavior
that is consistent with the quintom model which allows $w$ to cross
-1. The transition redshift at which $w$ departs from -1 is located
at $z\sim 2.0$, which might be overlooked by the previous analysis
fixing $w=-1$ at $z>1.5$. After adding constraints from the galaxy
cluster measurements ($0.142<z<0.890$), we provide a much tighter
limit on the EoS parameter: $w=-1.066^{+0.614}_{-0.614}$.
Considering the redshift coverage of these two astrophysical probes,
the combination of high-redshift quasars and low-redshift clusters
may provide an important source of angular diameter distances. In
order to asses the reliability of the above results with
intermediate-luminosity quasars, the effect of several systematics
on the final cosmological fits, due to the uncertainties of the
linear size scaling factor ($l$) as well as the dependence of $l_m$
on the luminosity and redshift ($\beta$, $n$), were also extensively
studied in our cosmological analysis. Our findings revealed that the
reduction of the above uncertainties will lead to more stringent
cosmological fits, which motivates the future use of VLBI
observations based on better uv-coverage to improve constraints on
$l$, $\beta$, and $n$ \citep{Pushkarev15}.

As a final remark, we point out that the sample discussed in this
paper is based on VLBI images observed with various antenna
configurations and techniques for image reconstruction. Our analysis
potentially suffers from this systematic bias, and taking it fully
into account will be included in our future work. To fully utilize
the potential of current and future VLBI surveys to constrain
cosmology, it will be necessary to reduce systematic errors
significantly. The largest current source of systematic uncertainty
is calibration. Calibration uncertainties can be split into
uncertainties related to the primary standard, and uncertainties in
the determination of the value of $l$, the linear size of this
standard rod. In principle, the first uncertainty can be reduced by
multi-frequency VLBI observations of more compact radio quasars with
higher sensitivity and angular resolution \citep{Cao17b}, while the
reduction of the second uncertainty should turn to more efficient
distance reconstruction technique. In this paper, we have applied
only one particular non-parametric method based on Gaussian
processes, to reconstruct angular diameter distances from 24 cosmic
chronometer measurements at $z \leq 1.2$. Application of new
distance-reconstruction techniques to future VLBI quasar
observations of high angular resolutions, will allow us to
cross-calibrate the quasar systems and significantly reduce the
systematic errors.

The QSO data set presented here and future complementary data sets
will help us to explore these possibilities. The approach,
introduced in this paper, would make it feasible to build a
significantly larger sample of standard rods at much higher
redshifts. With such a sample, we can further investigate
constraints on the cosmic evolution as well as possible evidence for
dynamical dark energy.

\section*{Acknowledgements}

Authors would like to thank the referee for careful and critical
reading and for valuable comments and suggestions which allowed to
improve our paper significantly. We are grateful to John Jackson for
his kind provision of the data used in this paper and for useful
discussions. We thank Zhengxiang Li and Meng Yao for helpful
discussions. This work was supported by the National Key Research
and Development Program of China under Grants No. 2017YFA0402603,
the Ministry of Science and Technology National Basic Science
Program (Project 973) under Grants No. 2014CB845806, the National
Natural Science Foundation of China under Grant Nos. 11503001,
11373014 and 11690023, the Fundamental Research Funds for the
Central Universities and Scientific Research Foundation of Beijing
Normal University, China Postdoctoral Science Foundation under Grant
No. 2015T80052, and the Opening Project of Key Laboratory of
Computational Astrophysics, National Astronomical Observatories,
Chinese Academy of Sciences. This research was also partly supported
by the Poland¨CChina Scientific \& Technological Cooperation
Committee Project No. 35-4. M.B. obtained approval of foreign talent
introducing project in China and gained special fund support of the
foreign knowledge introducing project.

\begin{table*}
\caption{\label{tab:data}Compilation of intermediate-luminosity
quasars : Column (1): source name; (2): angular size in
milliarcseconds; Column (3): uncertainty in angular size; Column
(4): total radio flux density at 2.29 GHz (Jy); Column (5): spectral
index; Column (6): optical counterpart (Q - Quasar; PQ - Probable
QSO).}

{\footnotesize\tiny
\begin{tabular}{lllllll|lllllllllll}

\hline\hline

Source & $z$ & $\theta$ &$\sigma_\theta$ & S & $\alpha$ & Type  & Source & $z$ & $\theta$ &$\sigma_\theta$ & S & $\alpha$ & Type\\

\hline

P 2351-006  &  0.462 &  2.743 & 0.027 & 2.49 & -0.1  & Q                                    & P 2326-477  &1.299 &2.07 &0.047 &2.48 &0 &Q \\
3C 279   &  0.5362 &  1.454 & 0.052  & 11.8 & 0.1  & Q                                  &P 0448-392  &1.302 &1.387 &0.064 &0.89 &0 &Q \\
P 0252-549   & 0.539  & 1.049 & 0.077 & 1.94  & 0.1  & Q                                &   P 1449-012  &1.319 &1.682 &0.064 &0.803 &-0.1 &Q \\
P 1136-13 &  0.558  &  1.974 & 0.048  & 3.4 & -0.3  & Q                                  &   P 2312-319  &1.323 &0.528 &0.135 &0.746 &-0.3 &Q \\
P 0403-13   &  0.574  &  2.399  &  0.077 &  4  & 0.1 &  Q                                &    P 1327-311 &1.326 &0.438 &0.863 &0.5 &0.1 &Q \\
P 0920-39   & 0.591 & 1.933 &  0.097  &  2.1  & -0.2  & Q                                  &  P 1615+29  &1.339 &1.608 &0.117 &0.62 &-0.2 &Q \\
3C 345 & 0.5928 & 1.525 &  0.041 & 7.6 & 0 & Q                                   & P 1831-711 &1.356 &1.46 &0.058 &1.364 &-0.2 &Q \\
GC 1104+16 & 0.632 &1.762 &0.15 &1.1 &-0.1 & Q                                 & P 0522-611  &1.4 &1.168 &0.096 &0.722 &-0.1 &Q \\
P 0113-118 &0.67 &2.218 &0.108 &2.7 &-0.1 &Q                                   & P 0005-239  &1.407 &0.758 &0.14 &0.59 &-0.1 &Q \\
P 2329-415  &0.6715 &1.292 &0.208 &1.1 &-0.1 &Q                                &    P 2320-035  &1.41 &0.596 &0.171 &0.79 &-0.1 &Q \\
GC 2251+13  &0.673 &1.972 &0.11 &1.1 &-0.3 &Q                                  & GC 0820+56  &1.417 &0.424 &0.114 &0.958 &-0.3 &Q \\
P 2344+09  &0.677 &1.904 &0.075 &1.9 &-0.2 &Q                                  & GC 0805+41  &1.42 &1.481 &0.306 &0.7 &-0.3 &Q \\
1803+78  &0.68 &1.244 &0.129 &2.6 &-0.1 &Q                                   & P 1532+01  &1.435 &1.172 &0.074 &1.19 &-0.3 &Q \\
0651+82  &0.71 &1.454 &0.245 &1.3 &0 &Q                                   & P 2335-18 &1.446 &1.81 &0.168 &0.725 &-0.3 &Q \\
P1354+19  &0.72 &1.415 &0.074 &1.8 &-0.1 &Q                                &  P 1030-357  &1.455 &1.718 &0.073 &0.682 &-0.2 &Q \\
GC 1636+47  &0.74 &1.139 &0.076 &1.06 &-0.1 &Q                              &    AO 0952+17 &1.472 &1.484 &0.223 &1 &-0.3 &Q \\
P 1237-10 &0.752 &1.518 &0.058 &1.63 &-0.2 &Q                               &     GC 2253+41 &1.476 &1.327 &0.19 &1.5 &-0.3 &Q \\
P 2135-248  &0.819 &0.897 &0.304 &0.7 &-0.2 &Q                               &  P 0524-460  &1.479 &1.784 &0.059 &0.895 &0.1 &Q \\
3C 179  &0.846 &1.953 &0.144 &1.7 &-0.3 &Q                                   & P 2052-47  &1.489 &1.292 &0.082 &1.05 &-0.3 &Q \\
P 0915-213  &0.847 &1.758 &0.094 &0.6 &-0.1 &Q                                 &  P 0220-349 &1.49 &1.243 &0.059 &0.6 &0 &Q \\
GC 1213+35 &0.857 &1.73 &0.183 &1.2 &-0.3 &Q                                   & P 2058-297 &1.492 &0.87 &0.429 &0.5 &-0.2 &Q \\
P 0454-46  &0.858 &2.094 &0.049 &2.439 &-0.2 &Q                                & 4C 46.29 &1.5586 &2.446 &0.216 &0.7 &0.1 &Q \\
P 1252+11  &0.871 &1.635 &0.249 &0.8 &-0.2 &Q                                   & P 2227-08  &1.5595 &1.038 &0.109 &1.3 &-0.1 &Q \\
P 1055+01 &0.888 &1.144 &0.059 &2.87 &0 &Q                                  & P 0406-127  &1.563 &0.796 &0.168 &0.58 &0.1 &Q \\
P 0537-441  &0.894 &1.149 &0.041 &3.777 &0.1 &Q                              &      P 0837+035  &1.57 &1.497 &0.1 &0.65 &-0.3 &Q \\
P 0537-158 &0.947 &1.747 &0.087 &0.64 &-0.1 &Q                                 &  P 1104-445  &1.598 &1.743 &0.081 &1.06 &0.1 &Q \\
P 2354-11  &0.96 &1.661 &0.121 &1.5 &-0.2 &Q                                   & P 1351+021  &1.6077 &1.788 &0.034 &0.347 &-0.3 &Q \\
P 1933-400  &0.965 &1.81 &0.063 &1.308 &0.1 &Q                                  &   P 0127+145  &1.6301 &1.788 &0.051 &0.579 &-0.2 &Q \\
GC 0237+04  &0.978 &0.568 &0.426 &0.8 &0.1 &Q                                 & P 1130+009  &1.633 &0.659 &0.203 &0.33 &0 &Q \\
P 0208-512  &0.999 &1.031 &0.051 &3.679 &-0.2 &Q                                &     P 0229-398  &1.646 &1.431 &0.093 &0.629 &0.1 &Q \\
P 0355-483  &1.016 &1.517 &0.033 &0.62 &-0.1 &Q                                 & NRAO 512  &1.66 &1.245 &0.214 &1.1 &0.1 &Q \\
P 0906+01  &1.018 &1.794 &0.081 &0.76 &-0.2 &Q                                & P 0922+005 &1.72 &1.005 &0.108 &0.94 &0 &Q \\
P 0130-17  &1.02 &1.544 &0.197 &1 &0 &Q                                 & P 0202-17  &1.74 &1.464 &0.124 &1.2 &0 &Q \\
OJ 320  &1.025 &1.552 &0.122 &1.2 &0 &Q                                &  P 1148-171  &1.751 &1.602 &0.205 &0.9 &-0.3 &Q \\
P 1348-289  &1.034 &1.797 &0.163 &1 &-0.3 &PQ                            &      DW 1403-08 &1.763 &1.629 &0.139 &0.73 &-0.3 &Q \\
P 2356+196  &1.066 &1.155 &0.453 &0.6 &0.1 &Q                             &      P 2320-021  &1.774 &1.25 &0.191 &0.33 &0 &Q \\
GC 1514+19 &1.07 &0.76 &0.07 &0.525 &0 &PQ                                 &   P 0108-079  &1.776 &0.66 &0.15 &1.054 &-0.2 &Q \\
P 0122-00  &1.08 &1.529 &0.16 &1.3 &-0.2 &Q                                 &   P 1451-400  &1.81 &1.136 &0.047 &0.734 &-0.2 &Q \\
GC 1144+40  &1.088 &1.399 &0.258 &0.9 &-0.2 &Q                              &      P 1034-374  &1.821 &0.518 &0.082 &0.567 &-0.3 &Q \\
GC 1335+55  &1.0987 &0.88 &0.581 &0.6 &-0.2 &Q                                &   P 0805-07 &1.837 &0.85 &0.495 &1.1 &0.1 &Q \\
P 2303-052  &1.139 &1.638 &0.086 &0.567 &-0.3 &Q                               &    0633+73  &1.85 &1.635 &0.307 &0.9 &-0.3 &Q \\
P 1210+134  &1.141 &1.727 &0.053 &0.514 &-0.1 &Q                                &  OK 492 &1.873 &1.13 &0.286 &1.1 &0.1 &Q \\
P 2329-16  &1.153 &1.758 &0.054 &1.2 &0.1 &Q                                    & OP-192  &1.89 &1.537 &0.059 &1.17 &0.1 &Q \\
P 1438-347  &1.159 &1.277 &0.069 &0.517 &-0.2 &Q                                 &   OH-230  &1.9 &1.396 &0.194 &0.7 &-0.2 &PQ \\
P 2332-017  &1.184 &1.485 &0.048 &0.57 &-0.3 &Q                                 & GC 1656+34  &1.939 &1.526 &0.163 &0.6 &-0.2 &Q \\
P 1127-14  &1.187 &1.611 &0.089 &0.79 &0 &Q                                    & P 0048-071  &1.975 &1.207 &0.094 &0.712 &-0.1 &Q \\
P 2329-384  &1.202 &1.099 &0.081 &0.796 &-0.2 &Q                                &    GC 0119+24 &2.025 &1.7 &0.279 &0.7 &0.1 &Q \\
P 1004-018  &1.212 &1.254 &0.088 &0.64 &0.1 &Q                                 &  OF 036  &2.048 &1.901 &0.148 &0.8 &-0.1 &Q \\
QC 08211+39 &1.216 &2.622 &0.085 &1.9 &-0.2 &Q                                  & P 0226-038  &2.055 &1.595 &0.057 &0.809 &-0.3 &Q \\
OV 591 &1.22 &2.032 &0.147 &1.4 &-0.1 &Q                                  & GC 1325+43  &2.073 &1.422 &0.368 &0.6 &-0.3 &PQ \\
P 1823-455  &1.244 &1.38 &0.064 &0.588 &-0.3 &Q                                 &   P 2319+07 &2.09 &1.916 &0.132 &0.9 &0 &Q \\
1150+81  &1.25 &2.44 &0.151 &1 &-0.1 &Q                                 & P 1116+12 &2.118 &0.564 &0.664 &0.5 &-0.3 &Q \\
VRO 40.09.02 &1.252 &2.109 &0.111 &1.7 &0 &Q                            &        P 2145-17  &2.13 &2.264 &0.048 &0.834 &-0.1 &PQ \\
GC 1020+40 &1.254 &1.544 &0.19 &1.2 &-0.3 &Q                             &       P 1020+191  &2.136 &1.449 &0.077 &0.57 &-0.3 &Q \\
GC 0537+53  &1.275 &1.046 &0.381 &0.8 &-0.3 &Q                             &       P 0642-349  &2.165 &1.097 &0.223 &1.2 &0.1 &Q \\
P 0514-16  &1.278 &1.883 &0.146 &0.7 &-0.1 &Q                               &   P 1032-199 &2.198 &1.434 &0.036 &1.082 &0.1 &Q \\
P 0405-385 &1.285 &0.979 &0.111 &2.2 &0.1 &Q                                &  P 2314-409  &2.448 &1.676 &0.045 &0.525 &-0.3 &Q \\
OR 186 &1.29 &1.583 &0.107 &0.87 &-0.3 &Q                                 & GC 1337+63  &2.5584 &1.598 &0.323 &0.6 &-0.2 &Q \\
GC 0707+47 &1.292 &0.517 &0.77 &0.8 &-0.3 &Q                               &     P 0329-255  &2.685 &1.506 &0.101 &0.417 &-0.1 &Q \\
P 0511-220 &1.296 &1.2 &0.187 &1.3 &0.1 &PQ                                 & P 0136+176  &2.73 &1.252 &0.132 &0.52 &0 &PQ \\

\hline\hline

\end{tabular}}

\end{table*}

\begin{table*}
\caption{ Measurements of the linear size of AGN compact structure
at different frequencies. ILQSO denotes Intermediate-Luminosity
Quasars. }
\begin{center} {\scriptsize
\begin{tabular}{|l|l|l|l|l|cc} \hline \hline
Frequency $\nu$  & Linear size $l (\nu)$  & Target systems  & Observational technique & Ref \\

\hline$\nu=2$ GHz & $l=11.03\pm0.25$ pc & 120 ILQSOs & VLBI & This paper  \\
\hline $\nu=4.6$ GHz & $l=5.9\pm5.0$ pc & Blazar 2200+420 & VLBA & \citet{OSullivan09} \\
\hline$\nu=5$ GHz & $l=5.59\pm0.07$ pc & ILQSO (J1611+0844) & VLBI & \citet{Frey10}  \\
\hline $\nu=5.1$ GHz & $l=5.4\pm4.5$ pc & Blazar 2200+420 & VLBA & \citet{OSullivan09} \\
\hline $\nu=7.9$ GHz & $l=2.5\pm0.7$ pc & Blazar 2200+420 & VLBA & \citet{OSullivan09} \\
\hline $\nu=8.9$ GHz & $l=2.5\pm0.6$ pc & Blazar 2200+420 & VLBA & \citet{OSullivan09} \\
\hline $\nu=12.9$ GHz & $l=1.9\pm0.4$ pc & Blazar 2200+420 & VLBA & \citet{OSullivan09} \\
\hline $\nu=15.4$ GHz & $l=1.6\pm0.3$ pc & Blazar 2200+420 & VLBA & \citet{OSullivan09} \\
\hline $\nu=22.2$ GHz & $l=1.1\pm0.3$ pc & Blazar 2200+420 & VLBA &\citet{OSullivan09} \\
\hline $\nu=43.1$ GHz & $l=0.5\pm0.1$ pc & Blazar 2200+420 & VLBA &\citet{OSullivan09} \\
\hline\hline
\end{tabular} }
\end{center}

\end{table*}

\begin{table*}

\caption{\label{constraint} Cosmological parameters $\Omega_m$,
$w_0$, $w_a$ and the Hubble constant $H_0$ fitted to the QSO data. }

 \begin{center}
{\scriptsize
 \begin{tabular}{|l|c|c|c|c|c|c|c|c|c|} \hline\hline
 Fit &                         $\Omega_m$        &     $w_0$    & $w_a$    &  $H_0 [\rm{km\;s}^{-1}\rm{Mpc}^{-1}] $    \\ \hline
 & \multicolumn{4}{c|}{ $\Lambda$CDM Model}  \\
  \cline{2-5}

QSO & $0.322^{+0.244}_{-0.141}$    & $-1$(fixed)     & $0$(fixed)     & $67.6^{+7.8}_{-7.4}$  \\
QSO(sys) & $0.312^{+0.295}_{-0.154} $     & $-1$(fixed)     & $0$(fixed)    & $67.0^{+11.2}_{-8.6}$ \\
QSO+CMB & $0.314^{+0.020}_{-0.020}$  & $-1$(fixed)  & 0(fixed)  & $68.76^{+1.95}_{-1.98}$  \\
QSO(Sys)+CMB  & $0.313^{+0.021}_{-0.020}$  & $-1$(fixed)  & 0(fixed)  & $68.87^{+4.65}_{-4.65}$ \\
QSO+CMB+BAO & $0.306^{+0.016}_{-0.014}$  & $-1$(fixed)  & 0(fixed)  & $69.45^{+1.58}_{-1.68}$  \\
QSO(Sys)+CMB+BAO  & $0.314^{+0.020}_{-0.018}$  & $-1$(fixed)  & 0(fixed)  & $68.79^{+4.58}_{-4.36}$ \\
QSO+CMB+BAO+Cluster & $0.306^{+0.016}_{-0.014}$  & $-1$(fixed)  & 0(fixed)  & $69.66^{+1.56}_{-1.58}$ \\
QSO(Sys)+CMB+BAO+Cluster  & $0.309^{+0.017}_{-0.015}$ & $-1$(fixed)  & 0(fixed)  & $69.04^{+2.79}_{-2.81}$ \\
\hline

 & \multicolumn{4}{c|}{ XCDM Model}  \\
 \cline{2-5}
QSO &$ 0.309^{+0.215}_{-0.151} $     & $-0.970^{+0.500}_{-1.730} $   & $0$(fixed)    & $67.3$(fixed) \\
QSO(sys)  &$0.295^{+0.213}_{-0.157}$    & $-1.130^{+0.630}_{-2.120}$    & $0$(fixed)   & $67.3$(fixed)\\
QSO+CMB & $0.329^{+0.023}_{-0.022}$  &$-0.941^{+0.078}_{-0.083}$  & $0$(fixed)    & $67.3$(fixed)  \\
QSO(Sys)+CMB & $0.323^{+0.057}_{-0.040}$ & $-0.956^{+0.183}_{-0.167}$ & $0$(fixed)  & $67.3$(fixed) \\
QSO+CMB+BAO & $0.331^{+0.022}_{-0.022}$ & $-0.939^{+0.075}_{-0.075}$  & $0$(fixed)    & $67.3$(fixed)  \\
QSO(Sys)+CMB+BAO & $0.331^{+0.042}_{-0.035}$ & $-0.937^{+0.139}_{-0.134}$ & $0$(fixed)  & $67.3$(fixed) \\
QSO+CMB+BAO+Cluster & $0.332^{+0.022}_{-0.022}$ & $-0.932^{+0.074}_{-0.075}$  & 0(fixed)  & $67.3$(fixed)  \\
QSO(Sys)+CMB+BAO+Cluster  & $0.335^{+0.042}_{-0.034}$ & $-0.916^{+0.131}_{-0.131}$ & 0(fixed) & $67.3$(fixed)  \\

\hline

 & \multicolumn{4}{c|}{ $w_z$CDM Model}  \\
 \cline{2-5}
QSO+CMB & $0.313^{+0.044}_{-0.024}$  & $-0.676^{+0.973}_{-0.707}$  & $-0.745^{+2.396}_{-5.501}$  & $67.3$(fixed) \\
QSO(Sys)+CMB  & $0.312^{+0.052}_{-0.038}$ & $-0.401^{+1.057}_{-0.992}$ & $-1.610^{+3.190}_{-6.371}$ & $67.3$(fixed)\\
QSO+CMB+BAO & $0.320^{+0.029}_{-0.022}$ & $-0.606^{+0.578}_{-0.544}$ & $-1.207^{+2.335}_{-2.684}$  & $67.3$(fixed) \\
QSO(Sys)+CMB+BAO  & $0.330^{+0.032}_{-0.035}$ & $-0.439^{+0.830}_{-0.681}$ & $-1.816^{+2.878}_{-3.493}$ & $67.3$(fixed) \\
QSO+CMB+BAO+Cluster & $0.320^{+0.025}_{-0.022}$  & $-0.532^{+0.488}_{-0.579}$  & $-1.686^{+2.424}_{-2.506}$ & $67.3$(fixed) \\
QSO(Sys)+CMB+BAO+Cluster  & $0.329^{+0.032}_{-0.033}$ &
$-0.375^{+0.637}_{-0.690}$ & $-2.330^{+2.868}_{-2.964}$ &
$67.3$(fixed) \\

 \hline\hline

 \end{tabular}}
 \end{center}

 \end{table*}

\begin{table*}
\caption{ Constraints on equation of state $w$ from the
redshift-divided radio quasar data. }
\begin{center} {\scriptsize
\begin{tabular}{|c|c|c|c|c|c|cccc}
\hline \hline EoS parameter & 0.46 $<$ z $<$ 1.0 & 1.0 $<$ z $<$ 1.5
& 1.5 $<$ z $<$ 2.0 & 2.0 $<$ z $<$ 2.5 &  2.5 $<$ z $<$ 2.73  \\
\hline $w[QSO]$ & $-0.99^{+0.25}_{-0.33}$ & $-0.92^{+0.20}_{-0.31}$
& $-1.52^{+0.53}_{-2.14}$ & $-0.30^{+0.13}_{-0.19}$ & $-1.08^{+0.70}_{-8.92} $ \\
\hline $w[QSO(sys)]$ & $-1.20^{+0.38}_{-0.56}$ & $-0.83^{+0.26}_{-0.45}$ & $-1.55^{+0.73}_{-5.79}$ & $-0.30^{+0.17}_{-0.25}$ & $-0.98^{+0.75}_{-9.02} $\\

\hline \hline
\end{tabular} }
\end{center}

\end{table*}

\begin{table*}
\caption{ Best fits for different cosmological models from the radio
quasar data. }
\begin{center} {\scriptsize
\begin{tabular}{|l|l|l|c} \hline \hline
Cosmological models  & Cosmological parameters & Cosmological parameters (sys)\\
\hline Flat cosmological constant &
$\Omega_m=0.322^{+0.244}_{-0.141}$, $H_0=67.6^{+7.8}_{-7.4}km/s/Mpc$
& $\Omega_m =
0.312^{+0.295}_{-0.154}$, $H_0=67.0^{+11.2}_{-8.6} \; km/s/Mpc$ \\
\hline Constant $w$  & ${\Omega_m}=0.309^{+0.215}_{-0.151}$,
$w=-0.97^{+0.50}_{-1.73}$ & ${\Omega_m}=0.295^{+0.213}_{-0.157}$,
$w=-1.13^{+0.63}_{-2.12}$ \\
\hline Ricci dark energy & ${\Omega_m}=0.229^{+0.184}_{-0.184}$,
$\beta=0.550^{+0.265}_{-0.265}$  &
${\Omega_m}=0.240^{+0.210}_{-0.210}$,
$\beta=0.520^{+0.365}_{-0.275}$ \\
\hline Dvali-Gabadadze-Porrati  &
$\Omega_m=0.285^{+0.255}_{-0.155}$, $H_0=66.2^{+7.4}_{-8.2}km/s/Mpc$
& $\Omega_m=0.248^{+0.335}_{-0.130}$, $H_0=64.3^{+11.8}_{-7.6}km/s/Mpc$\\
\hline \hline
\end{tabular} }
\end{center}

\end{table*}

\end{document}